\documentclass[12pt,reqno]{amsart}
\usepackage{amssymb,latexsym,amsmath,amstext}
\usepackage{amscd,amsfonts}
\usepackage{epsfig}

\newtheorem{theorem}{Theorem}

\theoremstyle{definition}
\newtheorem{remark}{Remark}

\numberwithin{equation}{section}

\begin{document}
\title{On the geometrical thermodynamics of chemical reactions}
\author{Manuel Santoro}\address{Department of Mathematics and Statistics, Portland State University, PO Box 751, Portland, OR
97207-0751, USA}
 \email{emanus@pdx.edu}
 \author{Albert S. Benight}\address{Department of Chemistry and Physics, Portland State University, PO Box 751, Portland, OR
97207-0751, USA}
 \email{abenight@pdx.edu}
\begin{abstract}
The formal structure of geometrical thermodynamics is reviewed with particular emphasis on the geometry of equilibria submanifolds.  On these submanifolds thermodynamic metrics are defined as the Hessian of thermodynamic potentials.  Links between geometry and thermodynamics are explored for single and multiple component, closed and open systems.  For single component, closed thermodynamic systems a detailed exposition is given which establishes a clear connection between the degeneracy and the scalar curvature of the Weinhold metric (Journal of Chemical Physcis, I-V, vol 63) and physical properties such as phase transition and non-ideal inter-particle interactions.  Compelling evidence for the relationship is provided by several specific applications. These include the Ideal gas, the van der Waals and the Berthelot gases.  For these cases, the degeneracy and the scalar curvature of the Weinhold metric are entirely consistent with the actual physical situation.  That is, for an Ideal gas no phase transition occurs and there are no interparticle interactions meanwhile the Weinhold metric is never degenerate and has zero scalar curvature. For the van der Waals and Berthelot gases, both display a phase transition and experience interparticle interactions, the Weinhold metric is generally degenerate along a sub-manifold of co-dimension one and has non-zero scalar curvature.
For multi-component closed and open systems the Gibbs free energy is employed as the thermodynamic potential to investigate the connection between geometry and thermodynamics.  The Gibbs free energy is chosen for the analysis of multicomponent systems and, in particular, chemical reactions. Major emphasis is focused on a detailed examination of single chemical reactions in a multi-component closed system. Then the approach is extended to consideration of $l$ independent chemical reactions.  For single chemical reactions, the general Gibbs metric for the Ideal gas mixture is provided.  Specific applications include the isothermal and isothermal-isobaric cases for a simple synthesis and single displacement chemical reactions.  For these simple systems, results suggest an intriguing relationship between non-ideal interparticle interactions and phase transitions.  Finally, the Gibbs metric is provided for multi-component open systems, including both ideal and non-ideal solutions.
\end{abstract}\bigskip
\maketitle
\section{Introduction}
\par
In the traditional approach to present the basic structure of
\textbf{homogeneous thermodynamics}, it is customary to fix a set
of variables describing the state of a system, the processes going
on in the system and interactions with the outside world. Such a
set usually includes the internal energy U of the thermodynamic
system. A set of values of these functions form the
\textbf{extended state space} of the system (referred to as the
energy-phase space in $[8]$, and thermodynamic phase space in
$[11]$). All but one of these variables, named thermodynamic
potential in this representation, are set into couples ($y$,$x$)
of intensive and extensive variables in such a way that to each
extensive variable $x$ there corresponds an intensive variable,
$y$, and the infinitesimal work (change of energy U or a chosen
thermodynamic potential) related to the change in the extensive
variable $x$ is,
\par
\[
dW=ydx
\]
\par
These couples are often collected into larger groups,
corresponding to the tensorial type of the process they describe
or to the process in whose description they participate.
Collecting all such pairs, the \textbf{first law of
thermodynamics} in its geometrical form postulates that during the
process, the change in the internal energy, $U$, is given by
integration over the trajectory in the state space of the one
form,
\par
\[
dU=dQ+\sum_{i}y_{i}dx^{i}
\]
\\
Here $dQ$ is the heat change one form. The \textbf{second law of
thermodynamics} in the formulation of C. Caratheodory [6,7] states
that the form,
\par
\[
dQ=dU-\sum_{i}y_{i}dx^{i}
\]
\\
has an integrating factor (see also [3]). After some effort [6,7],
this integrating factor is determined to be $\frac{1}{T}$ ($T$ is
the absolute temperature). Thus,
\par
\[
dQ=TdS
\]
\\
where the new state variable S is the entropy.
\par
The couple of variables ($S$,$T$), (entropy/temperature), plays a special
role in this formulation. If one lists all the extensive variables
$x_{i}$ (including entropy) and the corresponding intensive
variables $y_{i}$, the infinitesimal change of the internal energy of the
system is given by,
\par
\[
dU=\sum^{k}_{i=1}y_{i}dx^{i}
\]
\par
Thus, if the thermodynamic phase space in variables $(U;x^{i}, y_{i})$ is denoted as $P$,
there is a one-form $\theta$ defined by the choice of the process
and the variables related to it, namely,
\par
\begin{equation}\label{theta}
\theta=dU-\sum^{k}_{i=1}y_{i}dx^{i}
\end{equation}
\par
Processes that might occur in the system should be such that,
along the curve $t\rightarrow{r(t)}=(U(t),x^{i}(t),y_{i}(t))$,
$\theta(r'(t))=0$. Thus, in this geometrical situation the 1-form
$\theta$ defines the \textbf{contact structure} on $P$ and all the
physically admissible processes should be integrable curves of the
contact distribution $D_{p}=ker(\theta_{p})$, with $p\in{P}$, of
this structure [6,7].
\par
\subsection{Geometrical thermodynamics}
\par
Geometrical interpretations of equilibrium thermodynamics have
proved that state space is endowed with a canonical contact
structure that underlines the first law of thermodynamics.
Different representations of this structure in a \textbf{canonical
D'Arbois-chart} are related to different forms of the law of
conservation of energy which can be expressed through the internal
energy, entropy, Helmholtz free energy, or other extensive
variables [11,15]. Hermann [10] and later Mrugala [11]
\emph{argued that extended phase space of a homogeneous
thermodynamic system, endowed with the contact structure, is the
natural geometric space for descriptions of equilibrium
thermodynamics}. Until now, explicit application of this geometric
analysis to gain insight into, among other things, the critical
behavior of real chemical systems has not been presented.
Applications of this geometric approach to the analysis of simple
thermodynamic systems is the focus of the present study.
\par
To begin, R.Hermann [10] and later R.Mrugala [11] defined the
extended state space of a homogeneous thermodynamic system as a
\textbf{(2k+1)-dimensional manifold $P$ endowed with the contact
structure} given by a differential $1$-form $\theta$ such that
\par
\begin{equation}
\theta\wedge{(d\theta)^{k}}\neq{0}
\end{equation}
\\
where $\theta$ is called the contact form. This condition is
equivalent to the property of the smooth subbundle
$D\subset{T(P)}$ being as far from integrable as possible [1].
\par
Locally, the association $p\rightarrow{ker(\theta_{p})}$ defines a 2k-dimensional
distribution
\par
\[
p\rightarrow{D_{p}\subset{T_{p}(P)}}
\]
\\
independent of the choice of $\theta$.
\par
Moreover, it is possible to show that replacement of $\theta$ by
$f(p)\theta$, with some function $f(p)$ positive at all points of
domain of $\theta$, does not violate this condition. Any contact
form $\theta$, in an appropriate local canonical chart $\mathcal
C$ of variables $(x^{0},x^{i},y_{i},i=1,\ldots k)$, called the
D'Arbois chart, is expressed by [1],
\begin{equation}
\theta = dx^{0}-\sum_{i=1}^{k}y_{i}dx^{i}.
\end{equation}
\par
In such a canonical chart, the distribution D at the point $p$ is generated by the
vector fields,
\par
\begin{equation}
D_{p}=<\frac{\partial{}}{\partial{y_{i}}},y_{j}\frac{\partial{}}{\partial{x^{0}}}+\frac{\partial{}}{\partial{x^{j}}}>
\end{equation}
\\
and the differential of the form $\theta$, $d\theta$, defines on
each hyperplane, $D_{p}$, the symplectic structure
$(D_{p},d\theta)$. In local coordinates $d\theta$ has the
canonical form,
\par
\[
d\theta_{p}=\sum^{k}_{i=1}dx^{i}\wedge{dy_{i}}
\]
\\
Replacement of $\theta$ by $f(p)\theta$ leads to the replacement
of $d\theta$ by $df\wedge{\theta}+fd\theta$ which, after
restriction to a hyperplane $D_{p}$, becomes $fd\theta$. Thus,
contact structure alone defines only \textbf{conformally
symplectic structure} on the distribution D.
\par
On the manifold $P$ there exists a unique smooth vector field $Y$
called the \textbf{Reeb vector field} [5] such that,
\par
\begin{equation}
\theta(Y)=1\qquad \iota_{Y}(d\theta)=0
\end{equation}
\\
In particular one gets the canonical splitting,
\par
\begin{equation}
T_{p}(P)=D_{p}\oplus{\ker{d\theta_{p}}}
\end{equation}
\\
of the tangent bundle of $P$ into the direct sum of two
subbundles, the first being the subbundle of horizontal vectors of
the distribution D, while the second being the characteristic
subbundle of the form $d\theta$ [5]. Correspondingly, the
cotangent bundle $T^{*}(P)$ splits as well.
\par
In the canonical D'Arbois chart, the Reeb vector field is just,
\par
\[
Y=\frac{\partial{}}{\partial{x^{0}}}
\]
\par
Couples of variables $(x^{i},y_{i})$ denote pairs of independent
parameters ($x^{i}$) and corresponding conjugate variables
($y_{i}$) with respect to the chosen thermodynamic potential
$x^{0}$. Examples of $(x^{i},y_{i})$ pairs are: (1) temperature
and entropy $(T,S)$; (2) pressure and volume $(p,V)$; (3) mole
number of $i$-th component and corresponding chemical potential
$(N_{i},\mu _{i})$; extent of reaction and corresponding affinity $(\xi,A)$.
\par
\subsection{Thermodynamic Equilibrium}
\par
Thermodynamic equilibrium is a key notion in thermodynamics. In particular,
\\
\textit{in all systems there is a tendency to evolve toward states in which the properties are determined by intrinsic factors
and not by previously applied external influences. Such simple terminal states are, by definition,
 time independent. They are called equilibrium states} [4].
\par
In a state of thermodynamic equilibrium, intensive variables are functions of
the extensive variables, namely
\par
\begin{equation}
y_{i}=y_{i}(x^{j})
\end{equation}
\\
Then, choosing the internal energy $U$ as the potential, the form $\theta$ becomes
\par
\begin{equation}
\theta=dU-\sum^{k}_{i=1}y_{i}(x^{j})dx^{i}
\end{equation}
\par
Along the contact distribution $D=ker(\theta)$,
\par
\begin{equation}
dU=\sum_{i}y_{i}(x^{j})dx^{i}
\end{equation}
\par
Thus, the relation $y_{i}(x^{j})=\frac{\partial{U}}{\partial{x^{i}}}$ exists just on the \textit{maximal integral submanifold} of the
contact manifold $P$.
\par
Denoting $\Phi$ as a generic thermodynamic potential, it is known
that equilibrium states belong to a maximal integrable surface of
contact form $\theta$ in the space P determined by the choice of k
\emph{independent variables} $x^{i}$ and by the thermodynamic
potential $\Phi(x^{i})$ as a function of these variables
(constitutive relations). Another choice of independent variables
together with some other specification of $\Phi(x^{i})$ leads to
another equilibrium surface corresponding in general to another
constitutive relation. The core of the present study focuses on
{\bf Legendre submanifolds} defined to be maximal integral
k-dimensional submanifolds of $P$ on which the Pfaff equation
$\theta=0$ holds [1,11]. The standard approach to locally defining
such a submanifold, $\mathcal{S}_{\Phi}$, in terms of a generating
function, $\Phi$, is given by the following theorem:
\par
\begin{theorem}
(V.Arnold, [1]).
\par
For any partition $I\cup{J}$ of the set of indices (1,...,k) into
two disjoint subsets I,J and for a function $\Phi{(y_{I},x^{J})}$
of k variables $y_{i}$ with $i\in{I}$ and $x^{j}$ with $j\in{J}$,
the following equations,
\par
\begin{equation}
 x^{0}=\Phi-y_{i}\frac{\partial{\Phi}}{\partial{y_{i}}},
x^{i}=-\frac{\partial{\Phi}}{\partial{y_{i}}},
y_{j}=\frac{\partial{\Phi}}{\partial{x^{j}}}
\end{equation}
\\
define a Legendre submanifold $\mathcal{S}_{\Phi}$ of a contact
manifold $(P^{2k+1},D)$.
\par
Conversely, every Legendre submanifold of $(P^{2k+1},D)$, in a
neighborhood of any point, is defined by these equations for at
least one of $2^{k}$ possible choices of the subset I.
\end{theorem}
\par
In the special case in which $\Phi$ is a function
of only the independent variables ($x^{1}$,...,$x^{k}$), the Legendre submanifold (submanifold of equilibria states) $\mathcal{S}_{\Phi}$ is given by,
\begin{equation}
\mathcal{S}_{\Phi}=\{(\Phi,x^{i},y_{j},i=1,\ldots k)\in P\vert
\Phi=\Phi(x^{i}), y_{i}=\frac{\partial \Phi}{\partial
x^{i}},i\in{I})
\end{equation}
\par
On the integral submanifold $\mathcal{S}_{\Phi}$, the function
$\Phi{(x^{i})}$ can be defined in terms of the variables
$x^{i},y_{i}$ as,
\par
\begin{equation}
\Phi{(x^{1},...,x^{k})}=\sum_{i}x^{i}y_{i}(x^{1},...,x^{k})
\end{equation}
\par
In most cases, $(1.12)$ is homogeneous of order one and satisfies
the Euler equation, i.e.
\par
\begin{equation}
\Phi{(\lambda{x^{1}},...,\lambda{x^{k}})}=\lambda{\Phi{(x^{1},...x^{k})}}
\end{equation}
\\
(see [9]). The expression in $(1.12)$ leads directly to the
Gibbs-Duhem relation between variables along the integral
submanifold $\mathcal{S}_{\Phi}$. Indeed, taking the differential
of $\Phi$, we obtain
\par
\[
0=\theta|_{\mathcal{S}_{\Phi}}=d\Phi-\sum_{i}y_{i}dx^{i}=\sum_{i}x^{i}dy_{i}
\]

Thus, the basic contact condition $\theta=0$ is equivalent to
the Gibbs-Duhem constitutive relation,i.e.
\par
\[
\sum_{i}x^{i}dy_{i}=0
\]
\par
In order to apply the \textit{contact} formalism to equilibrium
thermodynamics, all variables $x^{0}$, $y_{i}$, $x^{i}$,
$i=1,...,k$, must be identified with the thermodynamic parameters
such that $\theta=dx^{0}-y_{i}dx^{i}$ satisfies the first law of
thermodynamics. These parameters \emph{only have physical meaning
on the k-dimensional Legendre submanifold} defined by the Pfaff
equation $\theta=0$. In this context, $x^{0}=\Phi$ is a
thermodynamic potential function of the independent variables
$x^{1}$,...$x^{k}$; and $y_{1}$,...,$y_{k}$ are the corresponding
conjugate parameters with respect to the potential. On the
Legendre submanifold, the parameters
$y_{i}=\frac{\partial{\Phi}}{\partial{x^{i}}}$ [13].
\par
\medskip
\par
One of the goals of the present study is, in the context of
geometrical thermodynamics, to show how the choice of extensive
variables and thermodynamic potential is crucial for a reasonable
physical interpretation of geometrical objects.
\par
\medskip
\par
In the present development, two sets of variables are chosen. In
the case of a \textbf{single component system}, $k=3$ and $x^{0}$
is identified with the internal energy, while, $x^{1}$, $x^{2}$,
$x^{3}$ are equated to the independent variables S, V, N,
respectively. Likewise $p_{1}$, $p_{2}$, $p_{3}$ are the
corresponding conjugate variables T, -p, $\mu$, respectively, and
the contact form becomes $\theta=dU-TdS+pdV-\mu{dN}$. This is an
extension of previous work [20] with the addition of a more
detailed exposition on the relation between geometry and
thermodynamics.
\par
In the case
of an \textbf{$r$-component system},
$k=r+2$ and $x^{0}$ is identified with the Gibbs free energy;
$x^{1}$, $x^{2}$, $x^{3}$,...,$x^{r+2}$ with the independent
variables T, p, $N_{1}$,...,$N_{r}$; and $p_{1}$, $p_{2}$,
$p_{3}$,...,$p_{r+2}$ with the conjugate variables -S, V,
$\mu_{1}$,...,$\mu_{r}$, respectively. Then, the contact form
becomes $\theta=dG+SdT-Vdp-\mu_{1}dN_{1}-...-\mu_{r}dN_{r}$.
\mathstrut
\par
\subsection{Thermodynamic Metric}
\par
A thermodynamic metric $\eta_{\Phi}$ defined by the constitutive
relation $\Phi=\Phi(x^{i})$ on the Legendre submanifold
$\mathcal{S}_{\Phi}$ of the contact structure $\theta$ has the
form([14,20]),
\begin{equation}
\eta _{\Phi}=\sum_{ij}\frac{\partial^{2} \Phi}{\partial
x^{i}\partial x^{j}}dx^{i}\otimes dx^{j}.
\end{equation}

For the case where $\Phi$ is the internal energy, $U$, the metric,
$\eta_{U}$, is called the Weinhold metric([21]). When $\Phi$ is
the entropy, $S$, the metric, $\eta_{S}$, is called the Ruppeiner
metric([17]). Here a new metric, $\eta_{G}$, is introduced for the
case where $\Phi$ is the Gibbs free energy, G.

\begin{remark} A thermodynamic metric, $\eta_{\Phi}$, of the form $(1.14)$ is induced
on $\mathcal{S}_{\Phi}$ by the following symmetrical tensor([20]),
\begin{equation}
{\tilde \eta}=\frac{1}{2}\sum_{i=1}^{k}(dy_{i}\otimes
dx^{i}+dx^{i}\otimes dy^{i}).
\end{equation}
\\
Up to a conformal factor, this tensor is the symmetrical tensor in
$P$ annihilating the Reeb vector field, $Y$, of structure
$\theta$, ($Y=\frac{\partial }{\partial x^{0}}$), and is invariant
under substitution of indices, $i$. ${\tilde \eta}$ is obtained as
the sum of symmetrical tensors in the 2-dimensional subspaces
$D_{i}^{*}$ of $D_{x}^{*} $ spanned by pairs of covectors
$(dx^{i},dy_{i})$ of thermodynamic conjugate variables [20].
\end{remark}

Moreover, thermodynamic metrics are generally degenerate and
non-definite. In particular the Legendre submanifold or (as it is
referred to hereafter) the \textbf{thermodynamic state space}, is
the union of domains where these metrics have different signatures
separated by the submanifold (generically of codimension one) of
states where these metrics are degenerate [20].
\par
\mathstrut
\par
\subsection{Scalar Curvature of the Thermodynamic Metric}
\par
Now consider a thermodynamic potential, $\Phi=\Phi(x^{i})$, a
function of the independent variables, $x^{i}$, and calculate the
scalar curvature of the corresponding thermodynamic metric defined
on the Legendre submanifold $\mathcal{S}_{\Phi}$. According to
expression $(1.14)$,
\par
\begin{equation}
\eta_{\Phi
ij}=\frac{\partial^{2}\Phi}{\partial{x^{i}}\partial{x^j}}.
\end{equation}
\\
Christoffel symbols for this metric are given by (see also [18]),
\par
\begin{equation}
\Gamma^{k}_{ij}=\frac{1}{2}\sum_{m}\eta_{ij,m}\eta^{km},
\end{equation}
where $\eta_{ij,m}=\frac{\partial \eta_{ij}}{\partial x^{m}}.$
\par
It can be shown that the curvature tensor of the metric
$\eta_{\Phi}$ is given by,
\par
\begin{equation}
\mathcal{R}^{l}_{ijk}=
\Gamma^{l}_{ki,j}-\Gamma^{l}_{ji,k}+\Gamma^{l}_{jp}\Gamma^{p}_{ki}-\Gamma^{l}_{kp}\Gamma^{p}_{ji}=
\frac{1}{4}(\eta_{ij,m}\eta_{sn,k}-\eta_{sn,j}\eta_{ki,m})\eta^{mn}\eta^{ls}
\end{equation}
\\
Therefore, the Ricci Tensor of metric $\eta_{\Phi}$ is given by,
\par
\begin{equation}
\mathcal{R}_{ik}=\mathcal{R}^{j}_{ijk}=\frac{1}{4}(\eta_{ij,m}\eta_{sn,k}-\eta_{sn,j}\eta_{ki,m})\eta^{mn}\eta^{js},
\end{equation}
\\
and the scalar curvature $\mathcal{R}_{\eta_{\Phi}}$ by (see also
[18]),
\par
\begin{equation}
\mathcal{R}_{\eta_{\Phi}}=\mathcal{R}_{ik}\eta^{ik}=\frac{1}{4}(\eta_{ij,m}\eta_{sn,k}-\eta_{sn,j}\eta_{ki,m})\eta^{mn}\eta^{js}\eta^{ik}.
\end{equation}
\par
\begin{remark}
Consider the scalar curvature of the metric defined on
two-dimensional integral surfaces $\mathcal{S}_{\Phi}$. Components
of the Ricci tensor are given by([20]),
\par
\[
\mathcal{R}_{11}=\frac{1}{4}(((\eta_{{21,}_{1}})^{2}-\eta_{{11,}_{1}}\eta_{{21,}_{2}})\eta^{11}\eta^{22}+(\eta_{{21,}_{1}}\eta_{{21,}_{2}}-\eta_{{11,}_{1}}\eta_{{22,}_{2}})\eta^{12}\eta^{22}+
\]
\par
\begin{equation}
((\eta_{{21,}_{2}})^{2}-\eta_{{11,}_{2}}\eta_{{22,}_{2}})(\eta^{22})^{2})
\end{equation}
\par
\[
\mathcal{R}_{12}=\mathcal{R}_{21}=\frac{1}{4}((\eta_{{11,}_{1}}\eta_{{21,}_{2}}-(\eta_{{21,}_{1}})^{2})\eta^{11}\eta^{12}+(\eta_{{11,}_{1}}\eta_{{22,}_{2}}-\eta_{{21,}_{2}}\eta_{{11,}_{2}})({\eta^{12}})^{2}+
\]
\par
\begin{equation}
(\eta_{{11,}_{2}}\eta_{{22,}_{2}}-(\eta_{{21,}_{2}})^{2})\eta^{12}\eta^{22})
\end{equation}
\par
\[
\mathcal{R}_{22}=\frac{1}{4}((\eta_{{21,}_{1}})^{2}-\eta_{{22,}_{1}}\eta_{{11,}_{1}})(\eta^{11})^{2}+(\eta_{{21,}_{2}}\eta_{{11,}_{2}}-\eta_{{11,}_{1}}\eta_{{22,}_{2}})\eta^{11}\eta^{12}+
\]
\par
\begin{equation}
((\eta_{{21,}_{2}})^{2}-\eta_{{12,}_{1}}\eta_{{22,}_{2}})\eta^{11}\eta^{22})
\end{equation}
\par

It is straight forward to show that,

\begin{equation}
\mathcal{R}_{11}\eta^{11}=\mathcal{R}_{22}\eta^{22}
\end{equation}

Thus, the scalar curvature is given by([20]),
\par
\begin{equation}
\mathcal{R}=2(\mathcal{R}_{11}\eta^{11}+\mathcal{R}_{12}\eta^{12})
\end{equation}
These general expressions for curvature are employed to
characterize the thermodynamics of several
systems.
\end{remark}
\par
\mathstrut
\par
\subsection{Closed and Open Thermodynamic Systems}
\par
Consider the thermodynamic state of a system as a function of a
certain number of independent variables such as the entropy, S,
volume, V, and number of moles, $N_{1}$,...,$N_{r}$, of r
components. Any function expressed in terms of these variables is
a \textit{state function} of the system. The First Law of
Thermodynamics or \textbf{Conservation of Energy} postulates the
existence of a state function, called the energy function, such
that the change in internal energy of the universe, given as the
sum of the energies of our system and of the surroundings, is
always constant, namely([16]),
\par
\begin{equation}
0=dU_{univ.}=dU_{sys.}+dU_{surr.}
\end{equation}
\\
This expression states that,
\par
\begin{equation}
dU_{sys.}=-dU_{surr.}
\end{equation}

The minus sign indicates a loss of energy by the surroundings.
Denoting $d_{E}U=-dU_{surr.}$ as the change in energy supplied to
the system by the surroundings, Conservation of Energy dictates,
\par
\begin{equation}
dU_{sys.}=dU=d_{E}U\qquad or\qquad d_{I}U=0
\end{equation}

Subscript $I$ indicates the energy change of the system. Note,
$d_{I}U=0$, is equivalent to $dU_{univ.}=0$.
\par
\mathstrut The Second Law of Thermodynamics or the principle of
\textbf{Entropy Production} postulates the existence of a state
function, called the entropy function, which possesses the
following properties: the entropy is an extensive variable, and
the change in entropy $dS$ can be separated into the flow of
entropy, $d_{E}S$, due to interactions with surroundings and a
term, $d_{I}S$, corresponding to entropy changes in the system
([16]). That is,
\par
\begin{equation}
dS=d_{E}S+d_{I}S
\end{equation}
\\
where $d_{I}S$ is denoted as the entropy production. $d_{I}S$ is
always non-negative, zero for reversible processes and positive
for irreversible ones.
\par
For \textit{closed} systems, conservation of energy in $(1.28)$
can be expressed as,
\par
\begin{equation}
dU=dQ+dW_{M}=dQ-pdV
\end{equation}
\\
with pressure, $p$, normal to the surface. For open systems,
\par
\begin{equation}
dU=d\Psi+dW_{M}=d\Psi-pdV
\end{equation}
\\
where $d\Psi$ is the infinitesimal rate of change of heat transfer
and exchange of matter of the system with the external environment
[16]. These expressions are employed to examine the geometries of
three different thermodynamic systems. First, work done on single
component thermodynamic systems is reviewed while stressing the
importance of choosing the right thermodynamic variables suitable
for a particular situation. For a one component thermodynamic
system physical interpretations are deduced from geometrical
objects such as the degeneracy and scalar curvature of the
Weinhold metric. Choosing the Gibbs free energy as the preferable
thermodynamic potential certain physical aspects of chemical
behaviour can be described through the geometry.  This approach is
also used in studying chemical reactions in multicomponent
systems. \mathstrut
\par
\subsection{System 1: Single Component Closed System}
\par
Consider a \emph{closed} system containing a single component in
the absence of an external field ([20]). The energy supplied by
the surroundings is derived from the sum of the heat transfer,
$dQ$, and mechanical work, $dW_{M}$. In this case the entropy
production $d_{I}S=0$, and the entropy of the system is given by,
\par
\begin{equation}
dS=d_{E}S=\frac{dQ}{T}
\end{equation}
\\
Therefore, in molar form the $1$-form of the energy, $du$, is
given by,
\par
\begin{equation}
du=dQ+dW_{M}=Tds-pdv
\end{equation}
\par
\mathstrut
\subsection{System 2: Multi-Component Closed System}
Consider a multi-component system in which changes in internal
energy can occur due to chemical reactions. In this case, the
entropy production $d_{I}S$ is given by [16],
\par
\[
d_{I}S=\frac{A}{T}d\xi
\]
\\
where A is the \textit{affinity} of the chemical reaction related
to the chemical potentials $\mu_{i}$ by
$A=-\sum_{i}\mu_{i}\nu_{i}$. The $\nu_{i}$ are the stoichiometric
coefficients and $\xi$ is the extent of reaction. Note, for a
single chemical reaction the entropy of the system is given by,
\par
\begin{equation}
dS=d_{E}S+d_{I}S=\frac{dQ}{T}+\frac{A}{T}d\xi
\end{equation}
\par
Now introduce another state function, the Gibbs free energy, G,
defined by,
\par
\[
G=U-TS+pV
\]
\\
In terms of this function, conservation of energy can be written
as [16],
\par
\begin{equation}
dG=dQ-TdS-SdT+Vdp=-SdT+Vdp-Ad\xi
\end{equation}
\par
While this expression is central to the present study, useful
geometrical tools are also introduced allowing a more general
treatment of the case of $l$ independent chemical reactions. In
this context, $(1.34)$ and $(1.35)$ can be restated as,
\par
\begin{equation}
dS=\frac{dQ}{T}+\sum_{n}\frac{A_{n}}{T}d\xi_{n},
\end{equation}
\\
and
\\
\begin{equation}
dG=-SdT+Vdp-\sum_{n}A_{n}d\xi_{n}
\end{equation}
\\
where $n=1,...,l$ is the number of chemical reactions that occur.
\par
\mathstrut
\par
\subsection{System 3: Open Systems}
Consider an open system in the absence of an external field
without entropy production. The expression in $(1.29)$ can be
generalized to consider changes in the number of moles
$N_{1}$,...,$N_{r}$ [16], viz.

\begin{equation}
dS=d_{E}S=\frac{d\Psi}{T}-\sum_{i}\frac{\mu_{i}}{T}dN_{i}
\end{equation}
\\
where $d\Psi=TdS+\sum_{i}\mu_{i}dN_{i}$ is the energy flow due to
heat transfer and exchange of matter. The corresponding Gibbs free
energy is given by,

\begin{equation}
dG=-SdT+Vdp+\sum_{i}\mu_{i}dN_{i}
\end{equation}
\\
with $i=1,...,r$, the number of moles of each reaction component.

\section{System 1: Single component system}
\par
Consider a $7$-dimensional thermodynamic phase space P of
variables\\ $(U,(S,T),(V,-p),(N,\mu))$ with the contact $1$-form given by,
\par
\begin{equation}
\theta=dU-TdS+pdV-\mu{dN}
\end{equation}
\\
where $N$ is the number of moles of the component and $\mu$ its
chemical potential. Next, consider a $3$-dimensional Legendre
submanifold $\mathcal{S}_{U}(S,V,N)$ of this system defined by the
constitutive relation,

\begin{equation}
U=U(S,V,N)
\end{equation}

By homogeneity of degree one of the internal energy, consider the
molar form of the constitutive relation $(2.2)$ and obtain the
following constitutive relation for a closed system with a single
component [4],
\par
\begin{equation}
u=u(s,v)
\end{equation}
\\
The differential is given by $(1.33)$, i.e.
\par
\[
du=Tds-pdv
\]
\\
Introduce the following thermodynamic parameters:
\par
\begin{enumerate}
\item $C_{V}$ is the heat capacity at constant volume:
\begin{equation}
C_{V}=T(\frac{\partial S}{\partial T})_{V},
\end{equation}
\item $C_{p}$ is the heat capacity at constant pressure:
\begin{equation}
C_{p}=T(\frac{\partial S}{\partial T})_{p},
\end{equation}
\item $\alpha$ is the thermal coefficient of expansion:
\begin{equation}
\alpha=\frac{1}{V}(\frac{\partial V}{\partial T})_{p}
\end{equation}
\item $k_{T}$ is the isothermal compressibility:
\begin{equation}
k_{T}=-\frac{1}{V}(\frac{\partial V}{\partial p})_{T}
\end{equation}
\end{enumerate}

For the molar case the above parameters are represented in lower
case type. The following expression relates the above molar
parameters:
\par
\begin{equation}
c_{p}-c_{v}=vT\frac{\alpha^{2}}{k_{T}}
\end{equation}
\par
Considering the expression in $(1.14)$, the Weinhold metric
$\eta_{u}$ is defined on the two-dimensional integral surface
$\mathcal{S}_{u}$ by [12],
\par
\begin{equation}
\eta_{ij_{u}}=\frac{\partial^{2}u}{\partial{x^{i}}\partial{x^{j}}}=\frac{1}{c_{v}}
\begin{pmatrix}
 T & -\frac{T\alpha}{k_{T}}\\
 -\frac{T\alpha}{k_{T}} & \frac{c_{p}}{vk_{T}}
\end{pmatrix}
\end{equation}
\par
It follows, if $T\neq{0}$, the Weinhold metric is degenerate along
the curve $\gamma_{\eta}$ given by,
\par
\begin{equation}
(\frac{\partial p}{\partial v})_{T}=0
\end{equation}
\\
which is presented in one of two forms: $p=p(v)$ or $T=T(v)$ [20].
\par
The main points of focus are as follows [20]:
\par
\textbf{1)} The equilibrium surface is the union of regions where the Weinhold metric has different signature separated by the curve $\gamma_{\eta}$ where the metric is degenerate;
\par
\textbf{2)} The critical point $(p_{c},T_{c},v_{c})$ of the system is the extremum of the functions $p=p(v)$ and $T=T(v)$;
\par
\textbf{3)} Along the curve of degeneracy, $\gamma_{\eta}$, a \emph{first order phase transition} seems to occur;
\par
\textbf{4)} Scalar curvature of the Weinhold metric is
\emph{strongly} influenced by parameters related to non-ideal
inter-particle interactions within the system.
\par
\begin{remark}
For a two dimensional state space, the determinant
$det(\eta_{\Phi})$ and the scalar curvature $\mathcal{R}$ of a
thermodynamic metric $\eta_{\Phi}$ are inversely related [20],

\begin{equation}
\mathcal{R}=-\frac{1}{4det(\eta_{\Phi} )^{2}}det  \begin{pmatrix}
\eta_{11} &
\eta_{11,1} & \eta_{11,2}\\
\eta_{12} &
\eta_{12,1} & \eta_{12,2}\\
\eta_{22} &
\eta_{22,1} & \eta_{22,2}\\
\end{pmatrix}.
\end{equation}
\end{remark}
In point 4) above, it was noted that the scalar curvature of the
Weinhold metric is strongly related to interparticle interactions
in the system. Moreover, as the determinant of the matrix
approaches zero, point 3) implies that the system approaches a phase transition at which point the scalar curvature $\mathcal{R}$ of the metric goes to infinity. Thus, the connection of points 3) and 4) suggests the following interpretation:
\par
\textbf{5)} If the system approaches a state "close enough" to the
curve of degeneracy, $\gamma_{\eta}$, the scalar curvature of the
metric goes to infinity.  Physically this is consistent with a
relevant increase in inter-particle interactions between the
reactant and product species when the system approaches a phase
transition.
\par
Such behavior suggests an intriguing relationship between
degeneracy, scalar curvature and inter-particle interactions. In
particular, this suggests a \emph{geometrical condition} for a
phase transition might be degeneracy (or infinite curvature) of
the Weinhold metric $\eta_{ij_{u}}$. The following examples
support this suggestion.
\medskip
\\
\textbf{Example I: Ideal gas.}
\\
Given the equation of state for an Ideal gas, $pv=RT$, the
Weinhold metric, $\eta_{ij_{u}}$, is given by [19],

\begin{equation}
\eta_{ij_{u}} =\frac{1}{c_{v}}
\begin{pmatrix}
 T & -p\\
 -p & \frac{c_{p}p}{v}
\end{pmatrix}
\end{equation}
\\
In this case $c_{p}-c_{v}=R$, and [20],
\par
\begin{equation}
det(\eta_{ij_{u}})=\frac{p}{vc^{2}_{v}}(Tc_{p}-pv)=\frac{pT}{vc_{v}}=\frac{RT^{2}}{c_{v}v^{2}}>0
\end{equation}
The above metric is positive definite on the costitutive surface
$\mathcal{S}_{u}$. Thus, for an Ideal gas, the energy metric is
never degenerate except for trivial cases, i.e. $T=0$. This lack
of degeneracy is consistent with the characteristics of an Ideal
Gas (i.e. it does not display a critical point and therefore does
not exhibit a phase transition). Moreover scalar curvature of the
Weinhold metric is zero, [17,20], i.e.
\par
\begin{equation}
\mathcal{R}_{\eta_{ij_{u}}}=0
\end{equation}
\par
Ruppeiner [17] was the first to suggest that zero curvature might
be evidence for the absence of inter-particle interactions. Their
absence is precisely the case for an ideal gas. Table $1$ gives
parallels between the geometric and thermodynamic features of an
Ideal Gas.
\medskip
\begin{center}
\par
\begin{table}[htbp]
\begin{tabular}{l|r} \hline
\textbf{Geometry} & \textbf{Thermodynamics}\\ \hline
No curve of degeneracy & No phase transition - no critical point\\ \hline
Zero scalar curvature & No inter-particle interaction\\ \hline
\end{tabular}
\medskip
\caption{Ideal Gas: Geometry-Thermodynamics}
\end{table}
\end{center}
\textbf{Example II: van der Waals gas}
\\
The equation of state for the van der Waals gas is given by,
\par
\begin{equation}
(p+\frac{a}{v^{2}})(v-b)=RT
\end{equation}
\\
where a and b are positive constants. Expression $(2.15)$ provides
a more realistic representation of the actual behavior of real
(non-ideal) gases by introducing the additional positive constants
a and b, characteristic of the particular gas under consideration.
The factor $(v-b)$ indicates the excluded volume of the molecules,
while the factor $\frac{a}{v^{2}}$ is the "interaction" term.
\par
For the van der Waals gas, the Weinhold metric is given by [20],
\par
\begin{equation}
\eta_{ij_{u}}=\begin{pmatrix}
 \frac{T}{c_{v}} & -\frac{TR}{(v-b)c_{v}}\\
 -\frac{TR}{(v-b)c_{v}} & (\frac{TR}{(v-b)^{2}}(1+\frac{R}{c_{v}})-\frac{2a}{v^{3}})
\end{pmatrix}
\end{equation}
\\
which is degenerate along the curve $\gamma_{\eta}$ written in the following forms:
\par
\[
    s=s(v)=c_{v}[(2+\frac{R}{c_{v}})\ln{(v-b)}+\ln{(\frac{2ac_{v}}{Rv^{3}})}]
\]
\par
\[
    p(v)=(v-2b)\frac{a}{v^{3}},\qquad\qquad\qquad T(v)=\frac{2a(v-b)^{2}}{Rv^{3}}
\]
\\
Note, in the limit $a=b=0$, the ideal case is recovered with the
metric in $(2.12)$. Taking the derivatives of the last two
expressions $p=p(v)$ and $T=T(v)$ and setting them to zero, the
critical point is obtained,
\par
\begin{equation}
(p_{c},T_{c},v_{c})=(\frac{a}{27b^{2}},\frac{8a}{27bR},3b)
\end{equation}
\par
Moreover, for the van der Waals gas the scalar curvature of the
Weinhold metric $\mathcal{R}_{\eta_{ij_{u}}}$ is given by [20],
\par
\begin{equation}
\mathcal{R}_{\eta_{ij_{u}}}=\frac{aRv^{3}}{c_{v}(pv^{3}-av+2ab)^{2}}
\end{equation}
\par

In general, the scalar curvature
$\mathcal{R}_{\eta_{ij_{u}}}\rightarrow{0}$ as $a\rightarrow{0}$
and as the system approaches the degeneracy curve,
$\mathcal{R}_{\eta_{ij_{u}}}\rightarrow{\infty}$. On the other
hand, expression $(2.18)$ does not vanish if the parameter
$b\rightarrow{0}$. So while the scalar curvature of the Weinhold
metric is not related to the excluded volume, it is strongly
influenced by the parameter $a$, that includes non-ideal
interactions in the system. This finding is entirely consistent
with the physical behavior of the van der Waals gas which exhibits
both a critical point and phase transition. Indeed in the $(p-T)$
plane, the following solutions are obtained [20],
\par
    \[
 p^{i}_{r}({T})=\frac{3v^{i}_{r}({T})-2}{(v^{i}_{r}({T}))^{3}}\qquad i=1,...,3
    \]
\\
One of the these solutions is the \textbf{Pressure-Temperature
Phase Boundary} (Fig.$1$).
 \par
\begin{figure}
  \includegraphics[height=2.0in, width=3.0in]{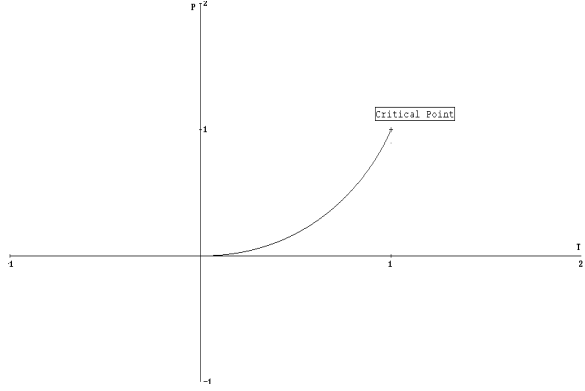}\\
  \caption{p-T phase boundary}\label{fig:clausius}
\end{figure}
\par
Table $2$ summarizes parallels between geometric and thermodynamic
features of the Van der Waals gas.
\medskip
\begin{center}
\par
\begin{table}[htbp]
\begin{tabular}{l|r} \hline
\textbf{Geometry} & \textbf{Thermodynamics}\\ \hline
Curve of degeneracy $\gamma_{\eta}$ & phase transition (see Fig.1)\\ \hline
Extremum of $\gamma_{\eta}$ & Critical point\\ \hline
Non-zero scalar curvature & Inter-particle interaction\\ \hline
\end{tabular}
\medskip
\caption{van der Waals: Geometry-Thermodynamics}
\end{table}
\end{center}
\medskip
\textbf{Example III: Berthelot gas}
\par
As a third example consider the Berthelot gas with the equation of
state,

\begin{equation}
(p+\frac{a}{Tv^{2}})(v-b)=RT
\end{equation}

where a and b are positive constants (analogous to the van der
Waals gas).
\par
The Weinhold metric for the Berthelot gas is given by [20],
\par
\begin{equation}
\eta_{ij_{u}} =
\begin{pmatrix}
 \frac{T}{c_{v}} & -\frac{1}{c_{v}}(\frac{RT}{(v-b)}+\frac{a}{Tv^{2}})\\
 -\frac{1}{c_{v}}(\frac{RT}{(v-b)}+\frac{a}{Tv^{2}}) &
 \frac{RT^{2}v^{3}-2a(v-b)^{2}}{Tv^{3}(v-b)^{2}}+\frac{T}{c_{v}}(\frac{R}{(v-b)}+\frac{a}{T^{2}v^{2}})^{2}
\end{pmatrix}
\end{equation}
\par
whose degeneracy is [20],
\par
\begin{equation}
RT^{2}v^{3}-2a(v-b)^{2}=2p^{2}v^{3}(v-b)^{2}-aR(v-2b)^{2}=0
\end{equation}
\par
In analogy to what was done for the van der Waals gas, it follows that
    \par
\[
    (v_{c},p_{c},T_{c})=(3b,\pm{(\frac{aR}{216b^{3}})^{\frac{1}{2}}},\pm{(\frac{8a}{27Rb})^{\frac{1}{2}}})
\]
\\
and that the scalar curvature of the Weinhold metric,
$\mathcal{R}_{\eta_{ij_{u}}}$, is given by [20],

\par
\begin{equation}
\mathcal{R}_{\eta_{ij_{u}}}=2a\frac{\left(T^{4}v^{4}Rc_{v}L(c_{v},v)+T^{2}v^{3}RaQ(c_{v},v)+a^{2}W(c_{v},v)\right)}{c^{3}_{v}T^{3}v(RT^{2}v^{3}-2a(v-b)^{2})^{2}},
\end{equation}
\\
where
\par
\[
L(c_{v},v)=(2c_{v}-R)v^{2}-3c_{v}bv+c_{v}b^{2}
\]
\par
\[
Q(c_{v},v)=-Rv^{5}+3Rbv^{4}-3Rb^{2}v^{3}+(Rb^{3}+c_{v}+R)v^{2}-b(b-2v)(R+c_{v})
\]
\\
and
\par
\[
W(c_{v},v)=-Rv^{7}+4Rbv^{6}-6Rb^{2}v^{5}+(2c_{v}+R+4Rb^{3})v^{4}-(8c_{v}+3R+Rb^{3})bv^{3}+(12c_{v}+3R)b^{2}v^{2}-
\]
\par
\[
-(8c_{v}+R)b^{3}v+2c_{v}b^{4}
\]
\par
The scalar curvature, $\mathcal{R}_{\eta_{ij_{u}}}$, goes to zero
as $a\rightarrow{0}$, and is strongly influenced by this parameter
which corresponds to non-ideal interactions in the system.
Furthermore, $\mathcal{R}_{\eta_{ij_{u}}}\rightarrow{\infty}$ as
the system approaches a phase transition. Once again, degeneracy
of the Weinhold metric and non-zero scalar curvature are
consistent with the characteristic physical behavior of the
Berthelot Gas. Parallels displayed in Table 2 for the van der
Waals gas are equally applicable to the Berthelot gas.
\par
\section{System 2: chemical reactions in a closed system}

\subsection{Single chemical reaction}
\par
Consider a closed system comprised of r components among which
chemical reactions can occur. First, we focus our attention on
single chemical reactions and then introduce multi-component
reactions. In a closed system, any change in the masses of the
components will occur only from a chemical reaction. Thus,
denoting the mass of component i by $m_{i}$, with $i=1,...,r$, the
infinitesimal change in mass can be written as [16],
\par
\begin{equation}
dm_{i}=\nu_{i}M_{i}d\xi
\end{equation}
\\
where $M_{i}$ is the molar mass of component $i$. The principle of
\textbf{conservation of mass} for a closed system is expressed as
[16],
\par
\begin{equation}
dm=\sum_{i}\nu_{i}M_{i}d\xi=0
\end{equation}
\\
with $m=\sum_{i}m_{i}$. The equation $\sum_{i}\nu_{i}M_{i}=0$ is
referred to as the \textbf{stoichiometric equation}.
\par
Alternatively, rather than the component masses it is more
convenient to consider the number of moles $N_{1}$,...,$N_{r}$
involved in the reaction. Since $\frac{dm_{i}}{M_{i}}=dN_{i}$, the infinitesimal change in the mole number of the $i$ component, can be expressed as
\par
\[
dN_{i}=\nu_{i}d\xi
\]
\\
Let $N^{0}_{i}$ be the number of moles of component $i$ in the
initial state of the system. When a reaction occurs, as indicated
by the stoichiometric coefficients $\nu_{i}$, the variations of
the number of moles of each component $N_{i}$ are not independent.
This can be expressed as [9],
\par
\begin{equation}
\frac{dN_{1}}{\nu_{1}}=...=\frac{dN_{i}}{\nu_{i}}=...=d\xi
\end{equation}
\\
where the \textbf{extent of the reaction}, $\xi$, is an extensive
variable just like the number of moles. Integrating and taking
$\xi=0$ as the initial state of the system, we obtain [9],
\par
\[
N_{i}=N^{0}_{i}+\nu_{i}\xi\qquad i=1,...,r
\]
\\
In this context, the Legendre submanifold $\mathcal{S}_{G}$ can be
defined by the constitutive relation, $G=G(T,p,\xi)$. Restriction of
the contact $1$-form, $\theta=dG+SdT-Vdp+Ad\xi$ to the submanifold
$\mathcal{S}_{G}$ provides,
\par
\begin{equation}
dG=-SdT+Vdp-Ad\xi
\end{equation}
\\
Thus, the general metric
$\eta_{ij_{G}}=\frac{\partial^{2}{G}}{\partial{x^{i}}\partial{x^{j}}}$,
where $x^{i}$ and $x^{j}$ are the extensive variables, is given
by,
\par
\begin{equation}
\eta_{ij_{G}}=
\begin{pmatrix}
 -\frac{C_{p}}{T} & {\alpha}V & -\Delta_{r}S\\
 {\alpha}V & -k_{T}V & \Delta_{r}V\\
 -\Delta_{r}S & \Delta_{r}V &
 -(\frac{\partial{A}}{\partial{\xi}})_{T,p}
\end{pmatrix}
\end{equation}
\\
where $-A=\Delta_{r}G$ is the \textbf{Gibbs free energy of the
reaction} and $\Delta_{r}S$ and $\Delta_{r}V$ are the entropy and
volume of reaction, respectively. Here, the affinity and Gibbs
free energy of reaction are used interchangebly.
\par
Naturally, as a reaction takes place, the chemical potential of
the components varies and so does the affinity of the reaction. At
constant temperature and pressure, the system is at equilibrium
whenever the affinity $A=0$. Since $(\frac{\partial{A}}{\partial{\xi}})_{T,p}\leq{0}$, the
determinant of the matrix in Eqn.$(3.5)$ is given by,
\par
\begin{equation}
\det{\eta_{ij_{G}}}=-\frac{C_{v}k_{T}V}{T}(\frac{\partial{A}}{\partial{\xi}})_{T,p}+\frac{C_{v}}{T}(\Delta_{r}V)^{2}+Vk_{T}[\Delta_{r}S-\frac{\alpha}{k_{T}}\Delta_{r}V]^{2}\geq{0}
\end{equation}
\par

\par
Of primary interest is what type of information is provided by the
degeneracy and, in some simple cases, by the scalar curvature of the metric
$\eta_{ij_{G}}$. As an example consider the three-dimensional
case of the Ideal gas mixture.
\medskip
\\
\textbf{Example I: Ideal gas mixture}.
\par
Consider a simple reaction in which substance A converts into
substance B. Starting with one mole of A, the relative amounts at
some later point in the reaction are $N_{A}=1-\xi$ and
$N_{B}=\xi$ with $\xi\in{[0,1]}$. Therefore, the Gibbs free energy can be written in
terms of the extent of reaction as,
\par
\[
G=(1-\xi)\mu_{A}+\xi\mu_{B}
\]
\par
Now, the chemical potential of an ideal gas mixture is given by
$\mu_{i}=\mu^{\theta}_{i}(T)+RT\ln({\frac{p_{i}}{p^{\theta}}})$
where $i=A,B$ and the superscript $\theta$ indicates some standard
state at pressure $p^{\theta}$. This reaction resulting in
conversion of the ideal component A into the ideal component B is
analogous to a homogeneous mixture of two ideal components, in
that the two components are mixed but do not interact.
\par
Considering that $p_{A}=(1-\xi)p$ and $p_{B}=\xi{p}$, where $p$ is
the total pressure, the Gibbs free energy can be written as,
\par
\[
G=(1-\xi)\mu^{\theta}_{A}+\xi\mu^{\theta}_{B}+RT\ln({\frac{p}{p^{\theta}}})+RT[(1-\xi)\ln({1-\xi})+\xi\ln{\xi}]
\]
\par
where $RT[(1-\xi)\ln({1-\xi})+\xi\ln{\xi}]=\triangle{G_{mix}}$ is
the \textit{Gibbs free-energy of mixing}.
\par
Thus, the metric $(3.5)$ becomes,
\par
\begin{equation}
\eta_{ij_{G}}=
\begin{pmatrix}
 (1-\xi)\frac{d^{2}\mu^{\theta}_{A}}{dT^{2}}+\xi\frac{d^{2}\mu^{\theta}_{B}}{dT^{2}} & \frac{R}{p} & \frac{d\mu^{\theta}_{B}}{dT}-\frac{d\mu^{\theta}_{A}}{dT}+
 R\ln(\frac{\xi}{1-\xi})\\
 \frac{R}{p} & -\frac{RT}{p^{2}} & 0\\
 \frac{d\mu^{\theta}_{B}}{dT}-\frac{d\mu^{\theta}_{A}}{dT}+
 R\ln(\frac{\xi}{1-\xi}) & 0 &
 \frac{RT}{\xi(1-\xi)}
\end{pmatrix}
\end{equation}
\\
where
$-\Delta_{r}S=\frac{d\mu^{\theta}_{B}}{dT}-\frac{d\mu^{\theta}_{A}}{dT}+
 R\ln(\frac{\xi}{1-\xi})$, $\Delta_{r}V=0$ and
 $(\frac{\partial{A}}{\partial{\xi}})_{T,p}=-\frac{RT}{\xi(1-\xi)}$.
The determinant in expression $(3.6)$ reduces to,
\par
\begin{equation}
\det{\eta_{ij_{G}}}=\frac{RT}{p^{2}}([\frac{d}{dT}(\mu^{\theta}_{B}-\mu^{\theta}_{A})+R\ln(\frac{\xi}{1-\xi})]^{2}-\frac{RT}{\xi(1-\xi)}[\frac{R}{T}+(1-\xi)\frac{d^{2}\mu^{\theta}_{A}}{dT^{2}}+\xi\frac{d^{2}\mu^{\theta}_{B}}{dT^{2}}])
\end{equation}
\par
If the chemical potentials of the two ideal components in the standard state are explicitly known, useful general information could be extrapolated from the Gibbs metric, its degeneracy and its scalar curvature. Since in general this is not the case, our analysis is restricted to the $2$-dimensional isothermal case and to the $1$-dimensional isothermal-isobaric case.
\medskip
\\
\textbf{{3.2 Isothermal single chemical reaction}.}
\par
When the temperature is kept constant during the
chemical reaction, $dT=0$ and the expression in Eqn. $(3.4)$
reduces to,
\par
\begin{equation}
dG=Vdp-Ad\xi
\end{equation}
\\
Thus, the metric $(3.5)$ reduces to,
\par
\begin{equation}
\eta_{ij_{G}}=
\begin{pmatrix}
 -k_{T}V & \Delta_{r}V\\
 \Delta_{r}V & -(\frac{\partial{A}}{\partial{\xi}})_{p}
\end{pmatrix}
\end{equation}
\\
with
\par
\begin{equation}
\det{\eta_{ij_{G}}}=k_{T}V(\frac{\partial{A}}{\partial{\xi}})_{p}-(\Delta_{r}V)^{2}
\end{equation}
\\
In the case of an ideal gas mixture, the metric of Eqn.$(3.10)$
becomes,
\par
\begin{equation}
\eta_{ij_{G}}=
\begin{pmatrix}
 -\frac{RT}{p^{2}} & 0\\
 0 & \frac{RT}{\xi(1-\xi)}
\end{pmatrix}
\end{equation}
\\
with determinant
$\det{\eta_{ij_{G}}}=-\frac{R^{2}T^{2}}{p^{2}\xi(1-\xi)}$ which is
always different than zero (except for trivial values of some
thermodynamic parameters). This implies that the Gibbs metric is,
in general, never degenerate for an Ideal mixture. Thus, the physical interpretation of this result is that there is
no \textit{critical} behavior displayed by an Ideal Gas Mixture. Moreover, the scalar curvature of the
metric $(3.12)$ is zero. Indeed, since the inverse of $\eta_{ij_{G}}$ is given by,
\par
\begin{equation}
\eta^{ij}_{G}=\begin{pmatrix}
 -\frac{p^{2}}{RT} & 0\\
 0 & \frac{\xi(1-\xi)}{RT}
\end{pmatrix}
\end{equation}
\\
the third derivatives of the Gibbs potential are
given by,
\par
\begin{equation}
\eta_{{11,}_{1}}=\frac{2RT}{p^{3}},\qquad
\eta_{{11,}_{2}}=\eta_{{12,}_{1}}=\eta_{{21,}_{1}}=0
\end{equation}
\par
\begin{equation}
\eta_{{22,}_{1}}=\eta_{{12,}_{2}}=\eta_{{21,}_{2}}=0,\qquad
\eta_{{22,}_{2}}=\frac{RT(2\xi-1)}{\xi^{2}(1-\xi)^{2}}
\end{equation}
\\
Therefore, the components of the Ricci tensor $\mathcal{R}_{ij}$,
$(1.21)$ to $(1.23)$, for an isothermal ideal mixture are all zero,
namely
\par
\[
\mathcal{R}_{ij}=0\qquad\ \forall i,j=1,2.
\]
\par
Using the expression in Eqn.$(1.25)$,
\par
\begin{equation}
\mathcal{R}_{\eta_{ij_{G}}}=0
\end{equation}
\par
Obviously, this result is consistent with the fact that the two \emph{Ideal} components when mixed do not interact, and it is essentialy consistent with the features of the single-component Ideal case. This strongly suggests that even in the context of chemical reactions in closed systems, non-zero scalar curvature might provide useful information regarding interactions between components. Although beyond the scope of the present study, this is an interesting path to pursue and, due to the complexity of the system, will require the use of numerical mathematics.
\medskip
\\
\textbf{3.3 Isothermal-isobaric single chemical reaction}
\par
It is interesting to note that in the case of constant temperature
and pressure, the change in Gibbs free energy is given by,
\par
\begin{equation}
dG=-Ad\xi
\end{equation}
\\
In this one-dimensional case, important information can be gleaned
from examination of the convexity of the Gibbs free
energy function.  Consider the condition,
\par
\begin{equation}
\frac{d^{2}G}{d\xi^{2}}=\frac{d\Delta_{r}G}{d\xi}=0
\end{equation}
\\
For an ideal gas mixture [9],
\par
\begin{equation}
\frac{d^{2}G}{d\xi^{2}}=\frac{RT}{\xi(1-\xi)}>{0},\qquad \xi\in{[0,1]},\qquad T\neq{0}
\end{equation}
\medskip
\\
\begin{figure}
  \includegraphics[height=4.0in, width=4.0in]{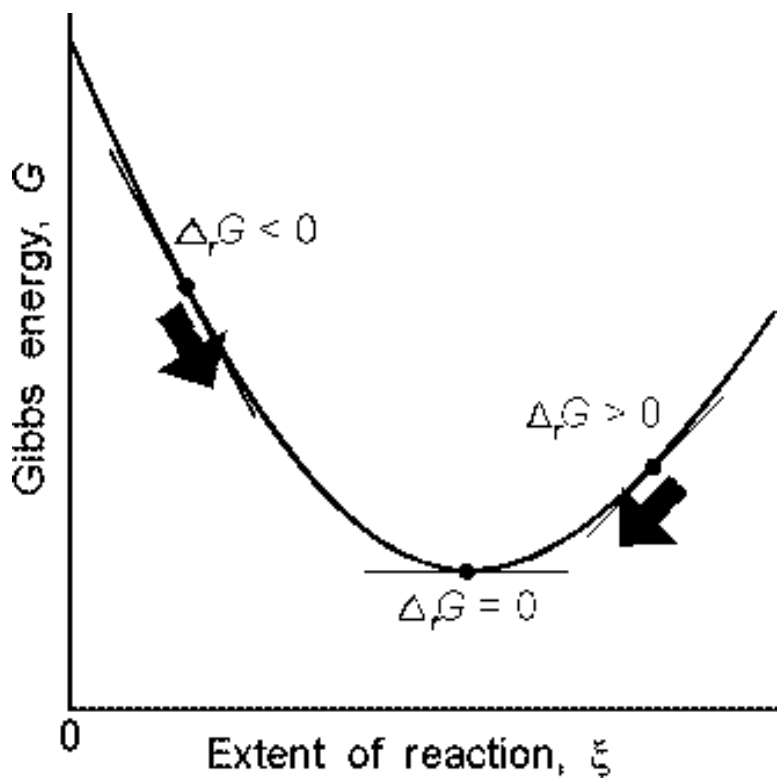}\\
  \caption{Ideal Gas Mixture: G vs. $\xi$}\label{fig:Gibbs}
\end{figure}
which implies that the Gibbs free energy is a convex function of
the extent of reaction. An example of such a function is displayed
in Fig.$2$, (see [22]), and defines the \textit{condition of
stability}. Initially, the Gibbs free energy decreases.  As a
reaction proceeds, the Gibbs free energy of the system continues
to decrease until it reaches a minimum value. At equilibrium
(constant T and p), the Gibbs energy is at the minimum, and
$\Delta_{r}G$ is equal to zero.  At greater extents of reaction
the Gibbs free energy is greater than zero and increases.
\par
This well-known result suggests for a non-ideal mixture the change
in sign and vanishing of the second derivative of the Gibbs
function might provide some insight into the stability of the
system. In particular, consider the case in which the Gibbs free
energy is not a simple convex function of the extent of the
reaction everywhere, but rather, first convex, then concave, then
convex again, as shown in Fig $3$. Points on the curve where
changes between convex and concave behavior occur are inflection
points which satisfy the expression in Eqn.$(3.18)$. Naturally, a
shift of the equilibrium state from one local minimum to another
constitutes a \textit{first order phase transition} induced by a
change in the extent of reaction [4]. For a chemical reaction, the
system tends to approach chemical equilibrium where the forward
and reverse reaction rates are the same and concentrations of the
reactant and product species do not change with time. When the
equilibrium condition is achieved, proportions of the various
compounds remain unchanged, and the reaction ceases to progress.
Prior to reaching the point of equilibrium, the system
\textit{fluctuates} between different equilibrium microstates.
Suppose that the system is confined in a lower (more stable) Gibbs
free-energy minimum and, occasionally, a fluctuation may be large
enough to push the system over the maximum to the region of higher
energy, i.e. a matastable minimum. A small fluctuation can
overcome the shallow barrier back to the more stable equilibrium
state [4]. Any thermodynamic system, in this case a chemical
reaction, tends to eventually reach the lowest minimum in the
Gibbs free energy. Naturally, if the "unstable" barrier is too
high or the minima are far apart a shift of the equilibrium from
one local minimum state to another is less probable.
\par
\begin{figure}
  \includegraphics[height=4.0in, width=4.5in]{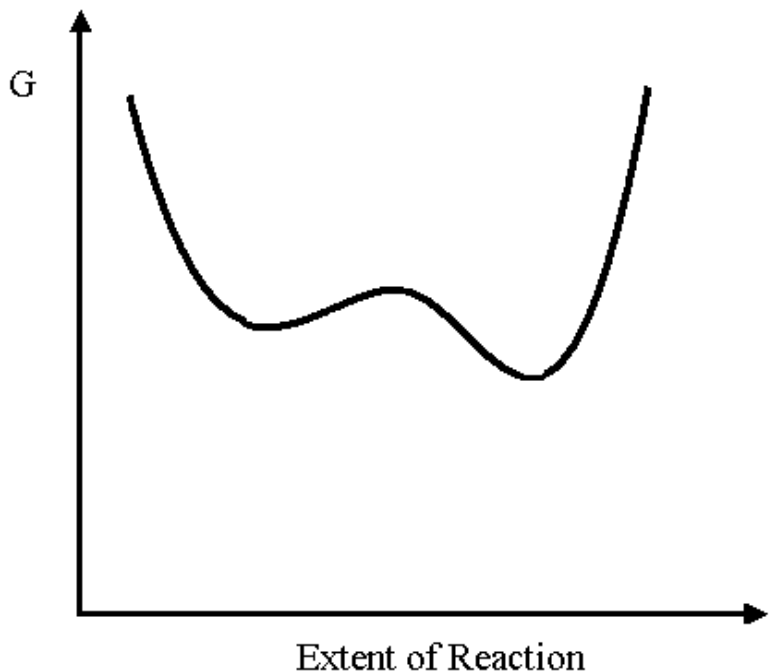}\\
  \caption{Non-ideal Mixture: G vs. $\xi$}\label{fig:Gibbs}
\end{figure}
\par
Within this picture, the local curvature of the Gibbs free energy
is positive for all points except those between the two inflection
points. Moreover, the portion of the curve between the minima at
the inflection points is said to be \textit{locally stable} but
\textit{globally unstable}. In this region on the curve metastable
states occur which appear to be stable to small perturbations, but
mixed configurations at the same extent of reaction represent more
stable states with lower free energy. A straight line connecting
the two minima corresponds to a phase boundary [4], i.e. a phase
transition from the phase at one minimum to the phase at the other
minimum. Positive local curvature fails at the points of
inflection. Local stability determines whether, after a small
perturbation, a system will return to the original equilibrium
state. Here our focus is on examining conditions leading to
failure of local stability.
\par
\begin{remark}
\textbf{Ideal Mixture}. For an isothermal and isobaric
ideal mixture,
\par
\begin{equation}
\frac{d\Delta_{r}G}{d\xi}=\frac{RT}{\xi(1-\xi)}
\end{equation}
\\
When $T\neq{0}$, consider that
\par
\begin{equation}
\frac{d\xi}{d\Delta_{r}G}=\frac{\xi(1-\xi)}{RT}=k\xi(1-\xi),\qquad
k=\frac{1}{RT}>0
\end{equation}
\\
which is the so-called \textbf{logistic equation} [2]. The process
described by this equation has two equilibrium positions, namely
$\xi=0$ and $\xi=1$.  Between these two points the field is
directed from $0$ to $1$. As a result the equilibrium position
$\xi=0$ is unstable (as soon as the reaction proceeds away from
$\xi=0$ reactants are converted to products). Meanwhile the
equilibrium position $\xi=1$ is stable. Moreover, integral curves
tend asymptotically to the line $\xi=1$ as
$\Delta_{r}G\rightarrow{+\infty}$ and to the line $\xi=0$ as
$\Delta_{r}G\rightarrow{-\infty}$. Such curves describe the
passage from one state (0) to another (1) in an \textit{infinite}
$\Delta_{r}G$.
\par
\end{remark}
\par
\subsection{Compounds}
\par
Consider a generic chemical reaction written as,
\par
\[
|\nu_{A}|A+|\nu_{B}|B+...\rightarrow{|\nu_{S}|S+|\nu_{T}|T+...}
\]
\\
where components on the left side are designated \emph{reactants}
while components on the right hand side are \emph{products} [9].
The $|\nu_{i}|$, $i=A,B,...$ are stoichiometric coefficients of
the reaction. Another formal representation of a chemical reaction
which better lends itself to mathematical manipulation is given by
[9],
\begin{equation}
\nu_{S}S+\nu_{T}T+...+\nu_{A}A+\nu_{B}B=\sum_{i}\nu_{i}i=0
\end{equation}
\\
The chemical potential of each component is given by [9],
\par
\begin{equation}
\mu_{i}=\mu^{*}_{i}(T,p)+RT\ln{\gamma_{i}x_{i}}=\mu^{*}_{i}(T,p)+RT\ln{a_{i}}\qquad
i=A,B,...
\end{equation}
\\
where $a_{i}$ is the \textit{activity} of component $i$. Recall
that $x_{i}=\frac{N_{i}}{\sum_{j}N_{j}}$, with
$N_{i}=N^{0}_{i}+\nu_{i}\xi$ and that $\gamma_{i}$ depends on the
extent of reaction. Thus, the Gibbs free-energy of the reaction is
given by [9],
\par
\begin{equation}
\Delta_{r}G=\sum_{i}\mu_{i}\nu_{i}=\sum_{i}\nu_{i}\mu^{*}_{i}+RT\ln{\prod_{i}a^{\nu_{i}}_{i}}=\sum_{i}\nu_{i}\mu^{*}_{i}+RT\ln{\prod_{i}\gamma^{\nu_{i}}_{i}x^{\nu_{i}}_{i}}
\end{equation}
\\
Introducing the \textit{quotient of reaction},
$Q_{a}=\prod_{i}a^{\nu_{i}}_{i}=\prod_{i}\gamma^{\nu_{i}}_{i}\prod_{i}x^{\nu_{i}}_{i}=Q_{\gamma}Q_{c}$,
where the subscript $c$ denotes concentration, the expression
$(3.24)$ can be written as [9],
\par
\begin{equation}
\Delta_{r}G=\Delta_{r}G^{\theta}+RT\ln{Q_{a}}
\end{equation}
\\
where
$\Delta_{r}G^{\theta}=\sum_{i}\nu_{i}\mu^{*}_{i}(T,p^{\theta})$ is
the standard Gibbs free energy of reaction.
It follows that,
\par
\begin{equation}
Q_{a}(\xi)=e^{\frac{\Delta_{r}G(\xi)-\Delta_{r}G^{\theta}}{RT}}
\end{equation}
\par
Naturally, if the system reaches equilibrium, namely
$\Delta_{r}G=0$, the parameter $Q_{a}$ is denoted by $K_{a}$, the
\textit{equilibrium constant}, and the expression in $(3.26)$
becomes [9],
\par
\begin{equation}
Q^{eq}_{a}=K_{a}=e^{-\frac{\Delta_{r}G^{\theta}}{RT}}
\end{equation}
\\
Thus, for an isothermal and isobaric single
chemical reaction,
\par
\[
\frac{d\Delta_{r}G}{d\xi}=RT\frac{d{\ln{Q_{a}}}}{d{\xi}}=RT[\frac{d{\ln{Q_{c}}}}{d{\xi}}+\frac{d{\ln{Q_{\gamma}}}}{d{\xi}}]
\]
\par
\begin{equation}
=RT[\sum_{i}(\frac{\nu^{2}_{i}}{N_{i}})-\frac{(\sum_{i}\nu_{i})^{2}}{\sum_{i}N_{i}}+\frac{d}{d{\xi}}\ln{(\prod_{i}\gamma^{\nu_{i}}_{i})}]
\end{equation}
\\
where
\par
\begin{equation}
\frac{d{\ln{Q_{c}}}}{d{\xi}}=\sum_{i}(\frac{\nu^{2}_{i}}{N_{i}})-\frac{(\sum_{i}\nu_{i})^{2}}{\sum_{i}N_{i}}
\end{equation}
\\
For simplicity, denote
$\frac{d}{d{\xi}}\ln{(\prod_{i}\gamma^{\nu_{i}}_{i})}=W(\xi)$.
Then, the expression in $(3.28)$ can be rewritten as,
\par
\begin{equation}
\frac{d^{2}G}{d\xi^{2}}=RT[\sum_{i}(\frac{\nu^{2}_{i}}{N_{i}})-\frac{(\sum_{i}\nu_{i})^{2}}{\sum_{i}N_{i}}+W(\xi)]
\end{equation}
\par
The expression in $(3.29)$ denotes \textit{the influence of the
relative amounts of reactants and products at each extent of the
reaction while $W(\xi)$ represents the relative strength of
non-ideal (inter-particle) interactions existent between products
and reactants}. Consequently, at any value of the extent of
reaction, there is a "possible" value of W such that the two
mentioned forces exactly balance one another. Thus, at a given
certain extent of the reaction determined by the relative amounts
of reactants and products, $W$ at that point corresponds to the
relative strength of non-ideal interactions that must exist
between products and reactants for a failure of local stability.
\par
\begin{remark}
For the ideal gas mixture, $Q_{a}=Q_{c}=\frac{\xi}{1-\xi}$ and therefore,
$\frac{dQ_{a}}{d\xi}=\frac{1}{(1-\xi)^{2}}>0$ (see Fig.
$4$).
In this case, $Q_{\gamma}=1$ and the expression in $(3.30)$ reduces to,
\par
\begin{equation}
\frac{d^{2}G}{d\xi^{2}}=RT[\sum_{i}(\frac{\nu^{2}_{i}}{N_{i}})-\frac{(\sum_{i}\nu_{i})^{2}}{\sum_{i}N_{i}}]
\end{equation}
\\
Moreover, when the sum of the stoichiometric coefficients vanishes
(i.e the isothermal-isobaric Ideal gas mixture, see $(3.19)$), the
expression in $(3.30)$ reduces further to,
\par
\begin{equation}
\frac{d^{2}G}{d\xi^{2}}=RT\sum_{i}(\frac{\nu^{2}_{i}}{N_{i}})>0
\end{equation}
\end{remark}
\par
\begin{figure}
  \includegraphics[height=6.0in, width=3.0in]{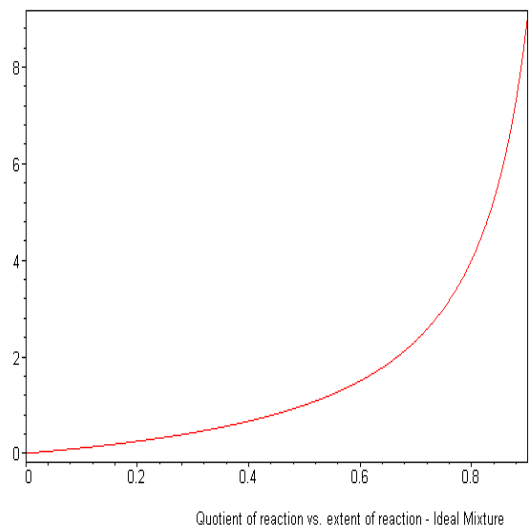}\\
  \caption{Ideal Mixture}\label{fig:ideal}
\end{figure}
\par
\begin{theorem}
Let $T\neq{0}$. Then, for an isobaric and isothermal single
chemical reaction,  $\frac{d^{2}G}{d\xi^{2}}=0$ if and only if
\par
\begin{equation}
W(\xi)=\frac{(\sum_{i}\nu_{i})^{2}}{\sum_{i}N_{i}(\xi)}-\sum_{i}\frac{\nu^{2}_{i}}{N_{i}(\xi)}
\end{equation}
\par
\end{theorem}
\par
The Gibbs free energy, $G$, is a convex function of
the extent of reaction whenever,
\par
\begin{equation}
W(\xi)>\frac{(\sum_{i}\nu_{i})^{2}}{\sum_{i}N_{i}(\xi)}-\sum_{i}\frac{\nu^{2}_{i}}{N_{i}(\xi)}
\end{equation}
\\
and a concave function whenever,
\par
\begin{equation}
W(\xi)<\frac{(\sum_{i}\nu_{i})^{2}}{\sum_{i}N_{i}(\xi)}-\sum_{i}\frac{\nu^{2}_{i}}{N_{i}(\xi)}
\end{equation}
\\
The curve described by the expression in $(3.33)$ is denoted as
the \textit{curve of phase boundary} in the $W$-$\xi$ plane. Such
a curve traces the phase boundary between the convex and concave
regions of the Gibbs free energy. In particular,  at a fixed value
of the extent of reaction, the system is locally stable whenever
the value of $W$ is less (in absolute value) than the value on the
curve of phase boundary. If instead the value of $W$ is greater,
the system is locally unstable.
\par
In early stages of the reaction, reactant species are far in
excess of product species, and $W$ is (in absolute value)
relatively large. Since a change from reactant phase to product
phase is improbable early in the reaction, interactions between
products, favoring product formation, must be greater than those
between reactants, favoring the reactant phase. $W$ indicates the
balance of the strengths of the product and reactant interactions
required for failure of local stability. As the reaction
progresses toward the extremum of the curve of phase boundary,
$\frac{dW}{d\xi} =0$, the relative difference in strength between
the two types of interactions is a minimum. At this critical
extent of reaction, a change between reactant and product phases
requires the smallest difference between their constituent
interactions and is thus most probable. Past this critical point
the extent of reaction increases.  To achieve local instability
the relative strength of the interactions favoring reactants must
be increasingly greater than those favoring the products.
\medskip
\\
\textbf{Example II: Synthesis Reaction}
\par
Consider a simple synthesis reaction in which two or more
substances combine to form a more complex substance. For example,
$2$ moles of di-hydrogen react with $1$ mole of oxygen to give $2$
moles of water,
\par
\begin{equation}
2H_{2}+O_{2}\rightarrow{2H_{2}O}
\end{equation}
\\
with
\par
\[
N_{H_{2}}=2-2\xi\qquad N_{O_{2}}=1-\xi\qquad and\qquad
N_{H_{2}O}=2\xi
\]
\\
and the corresponding stoichiometric coefficients given by,
\\
\[
\nu_{H_{2}}=-2\qquad \nu_{O_{2}}=-1\qquad and\qquad \nu_{H_{2}O}=2
\]
\\
Then, from $(3.33)$,
\par
\begin{equation}
W(\xi)=-\frac{6}{\xi(1-\xi)(3-\xi)}<0
\end{equation}
\\
A plot of the curve of phase boundary for this reaction is given
in Fig.5.
\par
\begin{figure}
  \includegraphics[height=6.0in, width=3.0in]{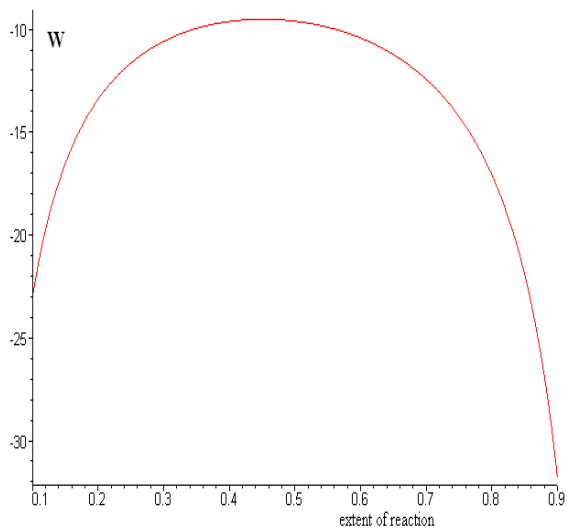}\\
  \caption{Synthesis Reaction: W vs.
$\xi$}\label{fig:ideal}
\end{figure}
\par
The extremum of this curve defines the "most probable" transition
point between the reactant and product phases.  At this point
differences in the relative amounts of reactant and product
species, and relative differences in the strengths of their
non-ideal interactions, are minimal. Taking the derivative with
respect to $\xi$ (Fig.$6$) and setting it to zero yields,
\par
\[
\xi=0.4514
\]
\par
\begin{figure}
  \includegraphics[height=6.0in, width=3.0in]{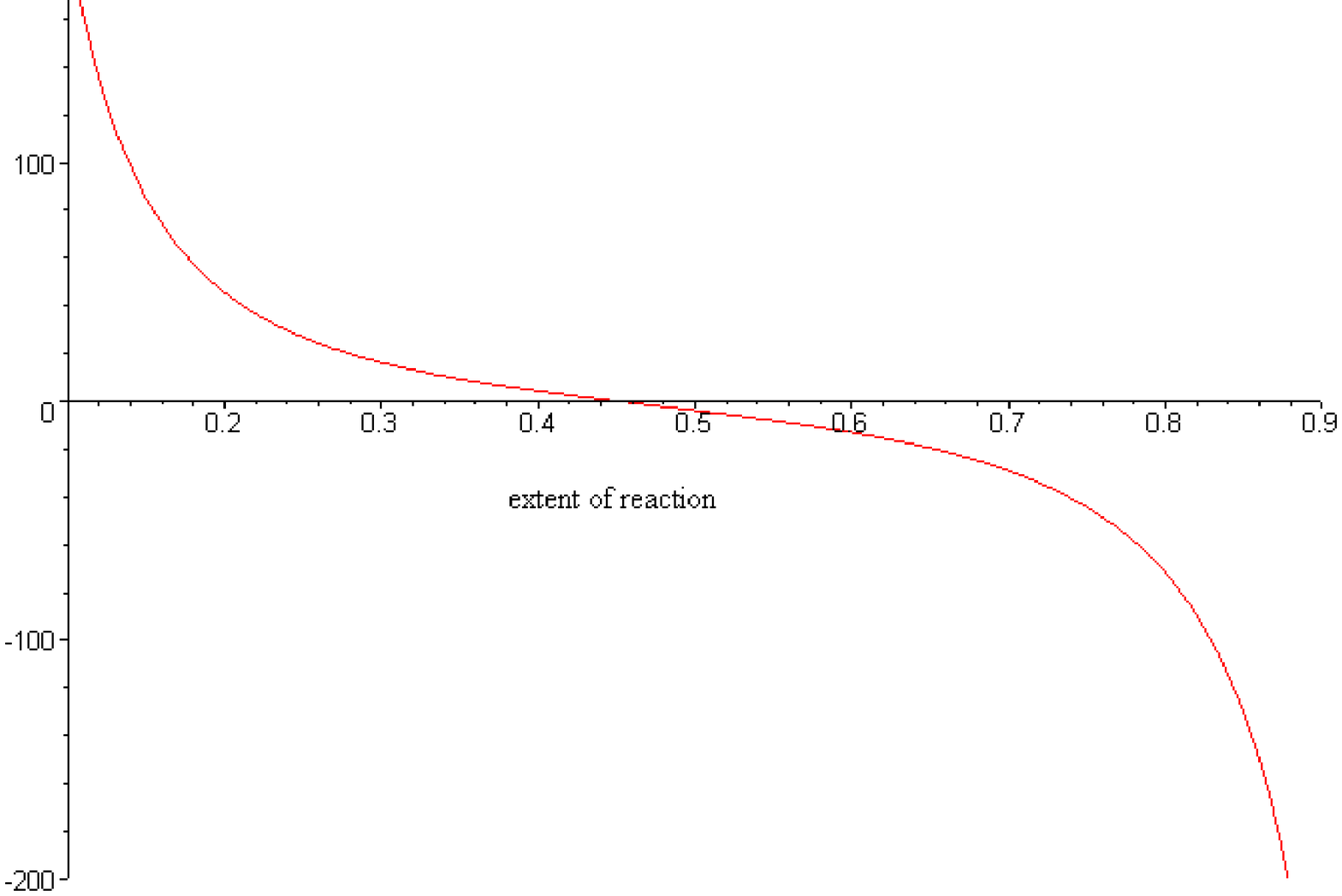}\\
  \caption{Synthesis Reaction: W' vs.
$\xi$}\label{fig:ideal}
\end{figure}
\par
It follows that $W=-9.5$. The fact that the critical extent of the
reaction is less than 0.5 suggests that the products and
associated non-ideal interactions are more strongly favored such
that the product phase is preferred even before half the extent of
reaction is reached.
\par
Now in analogy, consider the dissociation reaction, i.e. the
synthesis reaction in the opposite direction.  In this case $2$
moles of water split into $2$ moles of hydrogen and $1$ mole of
oxygen. Namely,
\par
\begin{equation}
2H_{2}O\rightarrow{2H_{2}+O_{2}}
\end{equation}
\par
Following the same steps as for analysis of the synthesis reaction,
the following expression for $W$ is obtained,
\par
\begin{equation}
W(\xi)=-\frac{6}{\xi(1-\xi)(2+\xi)}<0
\end{equation}
\\
The graph of this curve of phase boundary is shown in Fig.7.
\begin{figure}
  \includegraphics[height=6.0in, width=3.0in]{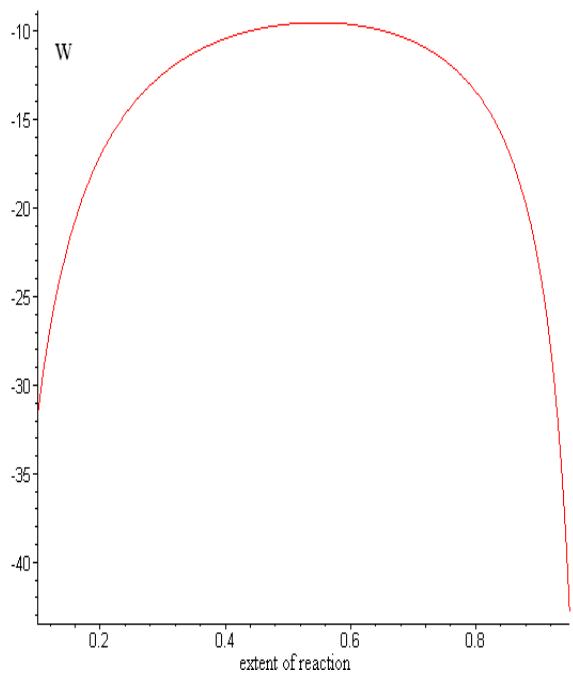}\\
  \caption{Dissociation Reaction: W vs.
$\xi$}\label{fig:ideal}
\end{figure}
Taking the derivative with respect to $\xi$ and setting it to zero
(see Fig.$8$),
\\
\[
\xi=0.5486
\]
\\
\begin{figure}
  \includegraphics[height=6.0in, width=3.0in]{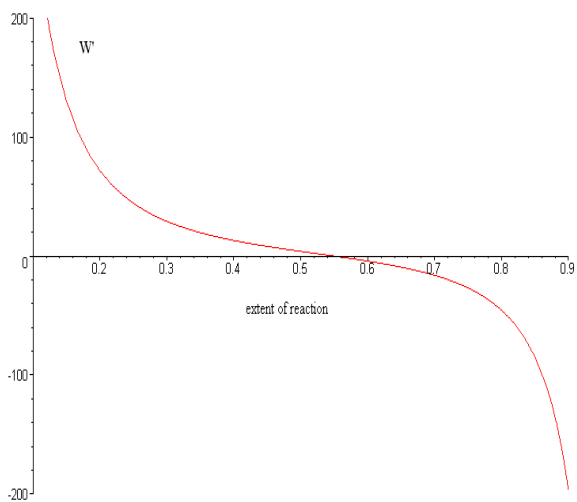}\\
  \caption{Dissociation Reaction: W' vs.
$\xi$}\label{fig:ideal}
\end{figure}
Note, $W=-9.5$ is apparently "invariant" to the direction of the
chemical reaction. The critical extent of reaction, $\xi$ = 0.5486
for the dissociation reaction, consistent with inferences drawn
from analysis of the synthesis reaction. That is, interactions
between the synthetic species ($H_{2}O$) are more favorable than
those between the individual species ($H_{2}$, $O_{2}$). The sum
of the two critical extents of reactions is $0.4514+0.5486=1$.
\medskip
\\
\textbf{Example III: Single Displacement Reaction}
\par
A single displacement reaction is one in which an atom (or ion) of
a single compound replaces an atom of another compound. As an
example, consider the single displacement in which copper ions in a
copper sulfate solution are displaced by zinc, forming zinc
sulfate:
\par
\begin{equation}
Zn+CuSO_{4}\rightarrow{Cu+ZnSO_{4}}
\end{equation}
\\
with,
\par
\[
N_{Zn}=1-\xi\qquad N_{CuSO_{4}}=1-\xi\qquad N_{Cu}=\xi\qquad
and\qquad N_{ZnSO_{4}}=\xi
\]
\\
The corresponding stoichiometric coefficients are given by,
\par
\[
\nu_{Zn}=-1\qquad \nu_{CuSO_{4}}=-1\qquad \nu_{Cu}=1\qquad
and\qquad \nu_{ZnSO_{4}}=1
\]
\\
Then, the expression in $(3.33)$ is given by,
\par
\begin{equation}
W(\xi)=-\frac{2}{\xi(1-\xi)}<0
\end{equation}
\\
The graph of this curve of phase boundary for this reaction is shown in Fig.9.
\begin{figure}
  \includegraphics[height=6.0in, width=3.0in]{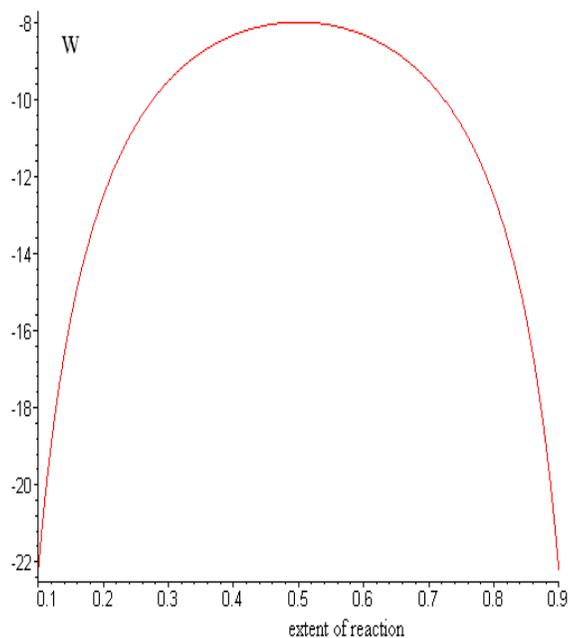}\\
  \caption{Single Displacement Reaction: W vs.
$\xi$}\label{fig:ideal}
\end{figure}
Taking the derivative with respect to $\xi$ and setting it to zero
(see Fig.$10$),
\par
\[
\xi=0.5
\]
\par
It follows that $W=-8$.  In this case $\xi=0.5$ corresponds to the
minimum difference between interactions associated with the
product and reactant species for a phase transition to occur. Note
that, for the displacement reaction, the critical extent of the
reaction is $0.5$. In this case, the number of products species
equals the number of reactants species and the critical extent of
the reaction is independent of the direction of the process. This
is in contrast to what was found for obtain in the synthesis and
dissociation reactions where product and reactant species are not
equal and the critical extent of the reaction depends on the
reaction direction.
\par
\begin{figure}
  \includegraphics[height=6.0in, width=3.0in]{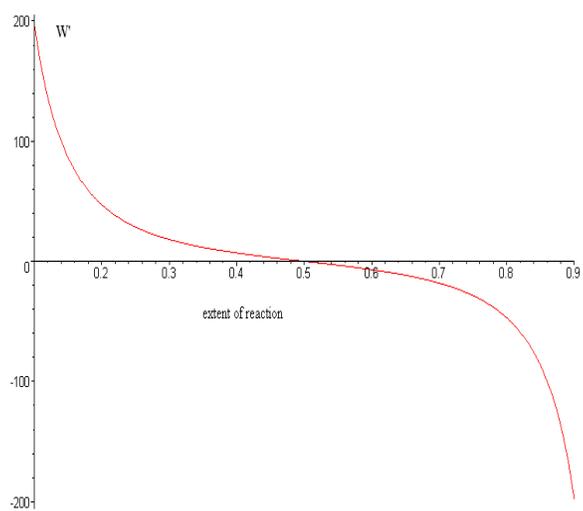}\\
  \caption{Single Displacement Reaction: W vs.
$\xi$}\label{fig:ideal}
\end{figure}
\par
\medskip
\par
\subsection{Multi-chemical reaction}
\par
Here the Gibbs metric is introduced for a closed system with r
chemical species in which $l$ independent reactions can occur. The
total change of mass, $dm_{i}$, is equal to the sum of the changes
resulting from the different reactions, [16],
\par
\begin{equation}
dm_{i}=M_{i}\sum_{n}\nu_{in}d\xi_{n}\qquad n=1,..,l
\end{equation}
\\
The principle of \textbf{Conservation of Mass} can be stated as
[16],
\par
\begin{equation}
dm=\sum_{i}\sum_{n}\nu_{in}M_{i}d\xi_{n}=0
\end{equation}
\\
As done previously, consider the number of moles of the components
in the system instead of their masses. The change in number of
moles of component $i$ is given by,
\par
\[
dN_{i}=\sum_{n}\nu_{in}d\xi_{n}
\]
\par
The Legendre submanifold $\mathcal{S}_{G}$ is defined by the constitutive
relation $G=G(T,p,\xi_{1},...\xi_{l})$ and the expression in
Eqn.$(3.3)$ becomes,
\par
\[
\frac{dN_{1,n}}{\nu_{1,n}}=...=\frac{dN_{i,n}}{\nu_{i,n}}=...=d\xi_{n}
\]
where $\xi_{n}$ is the extent of the n-th reaction with
$i=1,...,r$ and $n=1,...,l$, [9]. Now, consider the differential
of the Gibbs free-energy,
\par
\begin{equation}
dG=-SdT+VdP-\sum_{n}A_{n}d\xi_{n}
\end{equation}
\\
By recalling the expression in $(1.16)$, the following metric is
obtained,
\par
\begin{equation}
\eta_{ij_{G}}=
\begin{pmatrix}
 -\frac{C_{p}}{T} & {\alpha}V & -\Delta_{r}S_{1} & ... & -\Delta_{r}S_{r}\\
 {\alpha}V & -k_{T}V & \Delta_{r}V_{1} & ... & \Delta_{r}V_{r}\\
 -\Delta_{r}S_{1} & \Delta_{r}V_{1} &
 -(\frac{\partial{A_{1}}}{\partial{\xi_{1}}})_{T,p,\xi_{2},...\xi_{l}}
 & ... &
 -(\frac{\partial{A_{l}}}{\partial{\xi_{1}}})_{T,p,\xi_{2},...\xi_{l}}\\
 \vdots & \vdots & \vdots & ... & \vdots\\
 -\Delta_{r}S_{r} & \Delta_{r}V_{r} &
 -(\frac{\partial{A_{l}}}{\partial{\xi_{1}}})_{T,p,\xi_{2},...\xi_{l}} &... & -(\frac{\partial{A_{l}}}{\partial{\xi_{l}}})_{T,p,\xi_{1},...\xi_{l-1}}
\end{pmatrix}
\end{equation}
\par
This is the Gibbs metric for a multicomponent thermodynamic system in which $l$ independent chemical reactions occur.
\par
\section{System 3: Open systems}
\par
Finally, consider open systems in the absence of external fields.
Recall that the contact $1$-form $\theta$, restricted to the
Legendre submanifold $\mathcal{S}_{G}$ described by the
constitutive relation $G=G(T,p,N_{1},...,N_{r})$, gives the
following differential, see $(1.39)$,
\par
\begin{equation}
dG=-SdT+Vdp+\sum_{i}\mu_{i}dN_{i}
\end{equation}
\par
Define a \textit{partial molar quantity} as [9],
\par
\begin{equation}
\overline{X_{i}}=(\frac{\partial{X}}{\partial{N_{i}}})_{T,p,N_{j\neq{i}}}
\end{equation}
\\
Since the Gibbs metric is defined as the Hessian of the Gibbs potential,
$\eta_{ij_{G}}=\frac{\partial^{2}{G}}{\partial{x^{i}}\partial{x^{j}}}$,
where $x^{i}$ and $x^{j}$ are the extensive variables, the following result is obtained,
\par
\begin{equation}
\eta_{ij_{G}}=
\begin{pmatrix}
 -\frac{C_{p}}{T} & {\alpha}V & -\overline{S_{1}} & \cdots & -\overline{S_{r}}\\
 {\alpha}V & -k_{T}V & \overline{V_{1}} & \cdots & \overline{V_{r}}\\
 -\overline{S_{1}} & \overline{V_{1}} & \overline{\mu_{11}} &
 \cdots & \overline{\mu_{r1}}\\
 \vdots & \vdots & \vdots & \cdots & \vdots\\
 -\overline{S_{r}} & \overline{V_{r}} & \overline{\mu_{r1}} &
 \cdots & \overline{\mu_{rr}}
\end{pmatrix}
\end{equation}
\\
where
$\overline{\mu_{ik}}=(\frac{\partial{\mu_{i}}}{\partial{N_{k}}})_{T,p,N_{j\neq{k}}}$.
\par
At constant temperature and pressure, conservation of energy $(4.1)$ reduces to,
\par
\begin{equation}
dG=\sum_{i}\mu_{i}dN_{i}
\end{equation}
\\
Therefore, the metric in Eqn.$(4.3)$ becomes,
\par
\begin{equation}
\eta_{ij_{G}}=
\begin{pmatrix}
 \overline{\mu_{11}} & \overline{\mu_{21}} & \cdots & \overline{\mu_{r1}}\\
 \overline{\mu_{21}} & \overline{\mu_{22}} & \cdots & \overline{\mu_{r2}}\\
 \vdots & \vdots & \cdots & \vdots\\
 \overline{\mu_{r1}} & \overline{\mu_{r2}} & \cdots & \overline{\mu_{rr}}
\end{pmatrix}
\end{equation}
\par
\subsection{Ideal solutions}
\par
Now, introduce the Gibbs metric for the cases of an ideal and
non-ideal solution in an open thermodynamic system without
chemical reactions. A component $i$ in solution is said to be
\textit{ideal} when its chemical potential is given by, [9],
\par
\begin{equation}
\mu_{i}=\mu^{*}_{i}(T,p)+RT\ln{x_{i}}
\end{equation}
\\
where $\mu^{*}_{i}(T,p)$ is the standard state chemical potential
which is independent of the composition and
$x_{i}=\frac{N_{i}}{\sum_{j}N_{j}}=\frac{N_{i}}{N}$ is the mole
fraction. Note, the total number of moles, N, depends on the
number of moles of the r components, namely
$N=N(N_{1},...,N_{r})$.
\par
Using the expression $(4.6)$, the metric for an ideal
solution that depends on the temperature and the pressure of the
system is obtained,
\par
\begin{equation}
\eta_{ij_{G}}=
\begin{pmatrix}
 -\frac{C_{p}}{T} & {\alpha}V & (\frac{\partial{\mu^{*}_{1}}}{\partial{T}})_{p,N_{i}}+R\ln{x_{1}} & \cdots & (\frac{\partial{\mu^{*}_{r}}}{\partial{T}})_{p,N_{i}}+R\ln{x_{r}}\\
 {\alpha}V & -k_{T}V & (\frac{\partial{\mu^{*}_{1}}}{\partial{p}})_{T,N_{i}} & \cdots & (\frac{\partial{\mu^{*}_{r}}}{\partial{p}})_{T,N_{i}}\\
 (\frac{\partial{\mu^{*}_{1}}}{\partial{T}})_{p,N_{i}}+R\ln{x_{1}} & (\frac{\partial{\mu^{*}_{1}}}{\partial{p}})_{T,N_{i}} & RT(\frac{1}{N_{1}}-\frac{1}{N}) &
 \cdots & -\frac{RT}{N}\\
 \vdots & \vdots & \vdots & \cdots & \vdots\\
 (\frac{\partial{\mu^{*}_{r}}}{\partial{T}})_{p,N_{i}}+R\ln{x_{r}} & (\frac{\partial{\mu^{*}_{r}}}{\partial{p}})_{T,N_{i}} & -\frac{RT}{N} &
 \cdots & RT(\frac{1}{N_{r}}-\frac{1}{N})
\end{pmatrix}
\end{equation}
\\
For an \textbf{isothermal-isobaric} system such a metric becomes,
\par
\begin{equation}
\eta_{ij_{G}}=
\begin{pmatrix}
 RT(\frac{1}{N_{1}}-\frac{1}{N}) & -\frac{RT}{N} & \cdots & -\frac{RT}{N}\\
 -\frac{RT}{N} & RT(\frac{1}{N_{2}}-\frac{1}{N}) & \cdots &
 -\frac{RT}{N}\\
 \vdots & \vdots & \cdots & \vdots\\
 -\frac{RT}{N} & -\frac{RT}{N} & \cdots & RT(\frac{1}{N_{r}}-\frac{1}{N})
\end{pmatrix}
\end{equation}
\par
This metric is degenerate everywhere since $\det(\eta_{ij_{G}})=0$
for all values of the number of moles of the r components. This
result indicates the Gibbs free energy metric for an open system
does not distinguish differences between ideal components. As far
as the metric is concerned, due to the explicit lack of
inter-component interactions, the system is apparently comprised
of a single, indistinguishable, ideal component.
\par
\subsection{Non-Ideal solutions}
\par
For non-ideal systems, the chemical potential of component $i$
takes the form, [9],
\par
\begin{equation}
\mu_{i}=\mu^{*}_{i}(T,p)+RT\ln{\gamma_{i}x_{i}}
\end{equation}
\\
where $\gamma_{i}$ is the activity coefficient of species $i$ that
depends on temperature, pressure and composition of the solution.
Expression $(4.9)$ considers deviations from ideality.  The partial molar
entropy, $\overline{S_{i}}$, and the partial molar volume of component $i$,
$\overline{V_{i}}$, are given by,
\par
\begin{equation}
-\overline{S_{i}}=(\frac{\partial{\mu^{*}_{i}}}{\partial{T}})_{p,N_{i}}+R[\ln{\gamma_{i}x_{i}}+T(\frac{\partial{\ln{\gamma_{i}}}}{\partial{T}})_{p,N_{i}}]
\end{equation}
\\
and
\par
\begin{equation}
\overline{V_{i}}=(\frac{\partial{\mu^{*}_{i}}}{\partial{p}})_{T,N_{i}}+RT(\frac{\partial{\ln{\gamma_{i}}}}{\partial{p}})_{T,N_{i}}
\end{equation}
\\
Moreover,
\par
\begin{equation}
\overline{\mu_{ii}}=(\frac{\partial{\mu_{i}}}{\partial{N_{i}}})_{T,p,N_{j\neq{i}}}=RT[(\frac{1}{N_{i}}-\frac{1}{N})+(\frac{\partial{\ln{\gamma_{i}}}}{\partial{N_{i}}})_{T,p,N_{j\neq{i}}}]
\end{equation}
\\
and,
\par
\begin{equation}
\overline{\mu_{ik}}=(\frac{\partial{\mu_{i}}}{\partial{N_{k}}})_{T,p,N_{j\neq{k}}}=RT[-\frac{1}{N}+(\frac{\partial{\ln{\gamma_{i}}}}{\partial{N_{k}}})_{T,p,N_{j\neq{k}}}]
\end{equation}
\par
For symplicity, denote
$(\frac{\partial{\ln{\gamma_{i}}}}{\partial{N_{k}}})_{T,p,N_{j\neq{k}}}=\overline{(\ln{\gamma_{i}})}_{k}$
and retain the usual notation for partial molar entropy and
volume. Substituting the expressions $(4.10)$-$(4.13)$
into the metric $(4.3)$, the general metric of a multicomponent non-ideal
solution is obtained,
\par
\begin{equation}
\eta_{ij_{G}}=
\begin{pmatrix}
 -\frac{C_{p}}{T} & {\alpha}V & -\overline{S_{1}} & \cdots & -\overline{S_{r}}\\
 {\alpha}V & -k_{T}V & \overline{V_{1}} & \cdots & \overline{V_{r}}\\
 -\overline{S_{1}} & \overline{V_{1}} & RT[(\frac{1}{N_{1}}-\frac{1}{N})+\overline{(\ln{\gamma_{1}})}_{1}] &
 \cdots & RT[-\frac{1}{N}+\overline{(\ln{\gamma_{r}})}_{1}]\\
 \vdots & \vdots & \vdots & \cdots & \vdots\\
 -\overline{S_{r}} & \overline{V_{r}} & RT[-\frac{1}{N}+\overline{(\ln{\gamma_{r}})}_{1}] &
 \cdots & RT[(\frac{1}{N_{r}}-\frac{1}{N})+\overline{(\ln{\gamma_{r}})}_{r}]
\end{pmatrix}
\end{equation}
\\
and, in the case of an isobaric-isothermal non-ideal system, the Gibbs metric simplifies to,
\par
\begin{equation}
\eta_{ij_{G}}=RT
\begin{pmatrix}
 [(\frac{1}{N_{1}}-\frac{1}{N})+\overline{(\ln{\gamma_{1}})}_{1}] & [-\frac{1}{N}+\overline{(\ln{\gamma_{2}})}_{1}] & \cdots & [-\frac{1}{N}+\overline{(\ln{\gamma_{r}})}_{1}]\\
 [-\frac{1}{N}+\overline{(\ln{\gamma_{2}})}_{1}] &
 [(\frac{1}{N_{2}}-\frac{1}{N})+\overline{(\ln{\gamma_{2}})}_{2}]& \cdots & [-\frac{1}{N}+\overline{(\ln{\gamma_{r}})}_{2}]\\
 \vdots & \vdots & \vdots & \vdots\\
 [-\frac{1}{N}+\overline{(\ln{\gamma_{r}})}_{1}] & [-\frac{1}{N}+\overline{(\ln{\gamma_{r}})}_{2}] & \cdots & [(\frac{1}{N_{r}}-\frac{1}{N})+\overline{(\ln{\gamma_{r}})}_{r}]
\end{pmatrix}
\end{equation}
\par
\begin{equation}
=RT[\begin{pmatrix}
 (\frac{1}{N_{1}}-\frac{1}{N}) & -\frac{1}{N} & \cdots & -\frac{1}{N}\\
 -\frac{1}{N} & (\frac{1}{N_{2}}-\frac{1}{N}) & \cdots &
 -\frac{1}{N}\\
 \vdots & \vdots & \cdots & \vdots\\
 -\frac{1}{N} & -\frac{1}{N} & \cdots & (\frac{1}{N_{r}}-\frac{1}{N})
\end{pmatrix}+\begin{pmatrix}
 \overline{(\ln{\gamma_{1}})}_{1} & \overline{(\ln{\gamma_{2}})}_{1} & \cdots & \overline{(\ln{\gamma_{r}})}_{1}\\
 \overline{(\ln{\gamma_{2}})}_{1} & \overline{(\ln{\gamma_{2}})}_{2} & \cdots &
 \overline{(\ln{\gamma_{r}})}_{2}\\
 \vdots & \vdots & \cdots & \vdots\\
 \overline{(\ln{\gamma_{r}})}_{1} & \overline{(\ln{\gamma_{r}})}_{2} & \cdots & \overline{(\ln{\gamma_{r}})}_{r}
\end{pmatrix}]
\end{equation}
\par
\begin{equation}
=(\eta_{ij_{G}})_{ideal}+(\eta_{ij_{G}})_{deviation}
\end{equation}
\\
where the subscript "deviation" indicates the degree of
non-ideality.
\par
\section{Conclusions}
\par
Geometrical thermodynamics has been applied to study the behavior
of several simple thermodynamic systems.  For a single component,
closed system intriguing relationships between the geometrical
concepts of degeneracy and scalar curvature of the Weinhold
metric, and the physical concepts of a phase transition and
inter-particle, non-ideal interactions are divulged.  For
multi-component closed systems the Gibbs metric was presented and
the scalar curvature and degeneracy of this metric was determined
for a few simple cases and could be indicative of physical
behavior.  In summary, this study provides convincing examples
that this geometrical approach to analysis of thermodynamic
systems can be applied to actually divulge important physical and
chemical behavior.
\par


\begin{thebibliography}{99}
\bibitem{A} V. Arnold, \emph{Mathematical Methods of Classical Mechanics}, Springer Verlag.
\bibitem{A1} V. Arnold, \emph{Ordinary Differential Equations},
Springer Verlag.
\bibitem{JB} J. Boyling, \emph{Caratheodory's Principle and the Existence of Global Integrating Factors}, Commun.Math.Phys., 10, pp.52-68, 1968.
\bibitem{C} H.B. Callen, \emph{Thermodynamics},
    Whiley, 1960.
\bibitem{ACS} A. Cannas da Silva, \emph{Lectures on Symplectic Geometry}, Lecture Notes in Mathematics, Springer, 2000.
\bibitem{CA1} C. Caratheodory, \emph{Unterschungen uber die
Grundlagen der T, hermodynamyk}, Mathematishe Annalen, 67, pp.355-386,
1909.
\bibitem{CA2} C. Caratheodory, \emph{Unterschungen uber die
Grundlagen der Thermodynamyk}, Gesammelte Mathematische Werke,
B.2., Munchen, pp.131-177, 1955.
 \bibitem{G} J. Gibbs, \emph{The Collected Works}, v.1,
 Thermodynamics, Yale Univ. Press, 1948.
\bibitem{MGPI} M. Graetzel, P. Infelta, \emph{The Bases of
Chemical Thermodynamics}, Vol. I,II, Universal Publishers.
\bibitem{H} R. Hermann, \emph{Geometry, Physics and Systems},
    Dekker, N.Y., 1973.
\bibitem{M1} R. Mrugala, \emph{Geometrical Formulation of Equilibrium Phenomenological
Thermodynamics},Reports of Mathematical Physics, v.14,No.3,
pp.419-427, 1978.
\bibitem{M2} R. Mrugala, \emph{On equivalence of two metrics in classical
thermodynamics}, Physica A,v.125, pp.631-639, 1984.
\bibitem{M3} R. Mrugala, J.D. Nulton, J.C. Schoen, P. Salamon, \emph{Contact structure in thermodynamic
theory},  Reports of Mathematical Physics, v.29, No.1,
pp.109-121, 1991.
\bibitem{M4} R. Mrugala, \emph{On a Riemannian Metric on Contact Thermodynamic
Spaces}, Reports of Mathematical Physics, v.38, No.3,
pp.339-348, 1996.
\bibitem{P} S.Preston, \emph{Notes on the geometrical structures of homogeneous
thermodynamics}, manuscript, 2004.
\bibitem{IP} I. Prigogine, \emph{Thermodynamics of Irreversible
Processes}, Interscience Publishers, 1968.
\bibitem{R1} G.Ruppeiner, \emph{Thermodynamics:A Riemannian geometric model},
Phys. Review A. 20(4), pp.1608-1613, 1979.
\bibitem{R3} G. Ruppeiner,\emph{Riemannian geometry approach to
critical points: General theory}, Phys. Review E, v.57,n.5,
pp.5135-5145, 1998.
\bibitem{SNI} P.Salamon, J.Nulton, E. Ihrig,\emph{On the relation between entropy
and energy versions of thermodynamic length} J.Chem. Phys., 80,
436, 1984.
\bibitem{MP} M. Santoro, S. Preston,
    \emph{Curvature of the Weinhold Metric for Thermodynamical Systems with 2 degrees of freedom}, arXiv,org, math-ph/0505010, May 2005.
\bibitem{W} F. Weinhold, \emph{Metric Geometry of equilibrium
thermodynamics},p. I-V,Journal of Chemical Physics, v.63,
n.6,2479-2483, 2484-2487, 2488-2495,2496-2501,1976,
v.65,n.2,pp.559-564,1976.
\bibitem{INT} Fig.$2$ has been taken from "http://www2.mcdaniel.edu/Chemistry/ch307.notes/ Chemical20Equilibrium.html".
\end{thebibliography}
\end{document}